# On the nature of the two-positron bond: Evidence for a novel bond type

Mohammad Goli[1], Dario Bressanini[2] and Shant Shahbazian[3]

[1]School of Nano Science, Institute for Research in Fundamental Sciences (IPM), Tehran 19395-5531, Iran, E-mail: m_goli@ipm.ir

[2]Dipartimento di Scienza e Alta Tecnologia, Università dell'Insubria, Como, Italy
E-mail: dario.bressanini@uninsubria.it

[3]Department of Physics, Shahid Beheshti University, Evin, Tehran 19839-69411, Iran,
E-mail: sh_shahbazian@sbu.ac.ir

**Abstract**

The nature of the newly proposed two-positron bond in $(PsH)_2$, which is composed of two protons, four electrons and two positrons, is considered in this contribution. The study is done at the multi-component-Hartree-Fock (MC-HF) and the Diffusion Monte Carlo (DMC) levels of theory by comparing ab initio data, analyzing the spatial structure of the DMC wavefunction, and applying the multi-component quantum theory of atoms in molecules and the two-component interacting quantum atoms energy partitioning schemes to the MC-HF wavefunction. The analysis demonstrates that $(PsH)_2$ to a good approximation may be conceived of two slightly perturbed $PsH$ atoms, bonded through a two-positron bond. In contrast to the usual two-electron bonds, the positron exchange phenomenon is quite marginal in the considered two-positron bond. The dominant stabilizing mechanism of bonding is a novel type of classical electrostatic interaction between the positrons, which are mainly localized between nuclei, and the surrounding electrons. To emphasize its uniqueness, this mechanism of bonding is proposed to be called gluonic which has also been previously identified as the main deriving mechanism behind formation of the one-positron bond in $[H^-, e^+, H^-]$. We conclude that the studied two-positron bond should not be classified as a covalent bond and it must be seen as a brand-new type of bond, foreign to the electronic bonding modes discovered so far in the purely electronic systems.

## 1. Introduction

The concept of chemical bond is one of those handful of concepts without it modern chemistry could not be shaped and flourished.[1,2] From a theoretical viewpoint the analysis of quantum mechanical mechanisms behind the formation of chemical bonds as well as developing bond classification schemes are a century old, but yet an ongoing, mega-project with no consensus on its final outcomes.[3–11] Relegating the details of this mega-project, its central tenet is that electrons are the ultimate agents of bonding. This is exemplified in the Lewis-Langmuir model of bond, proposed even before the advent of quantum mechanics.[12–15] With such background, claiming novel bonding agents apart from the electron adds an unexpected and new dimension to the concept of chemical bond and goes beyond the "standard model" of bonding.

Accordingly, in a landmark computational study in 2018, Reyes and coworkers introduced the positron, the anti-matter analog of electron,[16] as the sole bonding agent between two hydrides, yielding a stable species denoted as $\left[ H^-, e^+, H^- \right]$.[17] Since then this claim was confirmed independently by high-level ab initio quantum chemical studies,[18,19] while new species bonded by a single positron were introduced as well.[20] More recently Bressanini has demonstrated computationally that apart from the aforementioned one-positron bonds, there are also two- and three-positron bonds,[21,22] opening a whole unexplored and rich area of bonding study. The positronic species are intrinsically unstable due to the electron-positron annihilation phenomenon, which is usually taking place in nanoseconds.[23–25] However, as is also the case for the purely electronic systems, the stability of the positronic species may also be considered relative to various dissociation channels. Whereas a positronic bond cannot prevent annihilation induced instability, it may stabilize a positronic species relative to some or all feasible dissociation channels. Demonstrating the stability of species bonded by positrons is a computationally non-trivial task since as will be discussed, various types of correlations must be taken properly into account in solving the Schrödinger equation for the positronic species.[26] An even more non-trivial issue is elucidating the nature of various discovered positronic bonds and answering questions relevant to their formation mechanisms. Are the mechanisms operative behind the positronic bond formation just like those forming the electronic bonds, or are there novel bond formation mechanisms yet to be discovered? Is it proper to classify these bonds based on the known categories of the electronic bonds, e.g. covalent, ionic, etc. or should new categories be invented?

In some primary studies to address these and similar questions there have been some attempts to elucidate the nature of the one-positron bond. Based on a comparative orbital-based analysis with the one-electron bonds in $H_2^+$ and $Li_2^+$, Reyes and coworkers originally concluded that the one-positron bond may be categorized as a type of covalent bond. In a subsequent and detailed study they considered the one-positron bonds in positron-bonded dihalides, $\left[ X^-, e^+, Y^- \right]$ ($X, Y = F, Cl, Br$).[20] Through an innovative energy partitioning scheme it was demonstrated that the positron binds the two halides through mainly stabilizing electrostatic interactions with the electrons of the two anions.[20] A comparative analysis with $\left[ A^+, e^-, B^+ \right]$ ($A, B = Na, K, Rb$) series revealed certain similarities between orbitals as well as the positronic and the electronic spin densities in these two series. Based on these similarities and relegating the observed stabilizing electrostatic interactions, they once again concluded that the one-positron bond deserved to be categorized as a covalent bond. In another study, Goli and Shahbazian employing the multi-component quantum theory of atoms in molecules (MC-QTAIM) partitioning methodology,[27–34] analyzed the AIM structure and their interactions in $\left[H^-,e^+,H^-\right]$.[35] According to their analysis, $\left[H^-,e^+,H^-\right]$ is indeed a diatomic species with two equivalent AIM, each containing a proton, two electrons, and half of the positron population. Interestingly, the used energy partitioning,[35–37] which is an extended version of the interacting quantum atoms (IQA) methodology developed for the purely electronic systems,[38–42] revealed that the interaction between AIM are mainly of the electrostatic nature. This type of electrostatic interaction is however quite distinct from the usual electrostatic interaction between oppositely charged ions where the charge imbalance, emerging from the electron migration between AIM, is the driving force behind the stabilizing interactions. The electrostatic interaction in $\left[H^-,e^+,H^-\right]$ originates from the stabilizing classical interaction of the positron density encompassed in one AIM with the electron density encompassed within the other AIM, whereas both AIM have equal overall charges, -0.5 in atomic units (a.u.). These results conform to the mentioned energy partitioning results obtained by Reyes and coworkers in their study on positronic dihalides and cast some doubt that the one-positron bond deserves to be categorized as a covalent bond. However, the term responsible for covalent bonding within the IQA partitioning scheme is the inter-AIM electronic exchange-correlation energy,[43,44] which only appears when there are two or more electrons in a system. As will be discussed in this paper, this is also true for the positronic bonds within the context of the extended IQA partitioning scheme, called the two-component IQA (TC-IQA). Thus, applying the TC-IQA to two- or three-positron bonds seems to be a better choice to consider whether the positron bonds may be categorized as covalent or not.

The first example of a two-positron bond was discovered by Bressanini through adding another positron to $\left[H^-, e^+, H^-\right]$, which was denoted as $(PsH)_2$.[21] $PsH$ is an atom made of a hydride ion and a positron and is usually conceived as a combination of hydrogen, $H$, and positronium, $Ps = e^- + e^+$, atoms.[45–48] Bressanini has proposed a duality scheme within which $H_2^+$ and $H_2$ are duals of $\left[H^-, e^+, H^-\right]$ and $(PsH)_2$, respectively.[22] Reyes and coworkers also proposed the first example of a multi-center two-positron bond through a detailed study of $2e^+\left[H_3^{-3}\right]$, which is a system composed of three hydride ions arranged in an equilateral triangle equilibrium geometry.[49] They did a detailed comparative study with $H_3^+$ and $Li_3^+$, as representatives of the two-electron three-center bonds. Their analysis revealed that $2e^+\left[H_3^{-3}\right]$ indeed possess a two-positron three-center bond but from a quantitative viewpoint, it was demonstrated that $2e^+\left[H_3^{-3}\right]$ is much more similar to $Li_3^+$ than $H_3^+$. This is also in line with their previously mentioned analysis where $\left[X^-, e^+, Y^-\right]$ and $\left[A^+, e^-, B^+\right]$ were proposed to be duals.[20]

Dualities of the hydrides containing positrons with lithium dimers or trimers are understandable considering $H^-$ is a polarizable ion like the lithium atom. Its static polarizability, $\sim 34\,\text{Å}^3$,[50] is much larger than that of the hydrogen atom, $\sim 0.7\,\text{Å}^3$,[50] but not very far from that of the lithium atom, $\sim 24\,\text{Å}^3$.[51] Interestingly, alkali dimers and their cations, i.e. $A_2$ and $A_2^+$ ($A = Li, Na, K, Rb, Cs$), in contrast to the deceptive simplicity of their molecular orbital structure,[52] are conceived of having anomalous bonds. This stems from the fact that the bond dissociation energies (BDEs) and bond lengths in the $A_2^+$ series, with a one-electron bond, are both concomitantly larger than their analogs in the $A_2$ series, with two-electron bonds.[53] These two quantities are usually inversely correlated and a longer bond length is associated with a smaller BDE and a weaker bond and vice versa. Various theoretical and computational studies have been conducted to explain this anomalous feature though it is hard to find a concrete consensus on a qualitative explanation in the relevant literature. It seems that the cause of this anomaly is an intricate interplay of the large polarization capacity of the alkali core electrons, the core-core interactions, and particularly, the strong valence-core correlations.[53–59] Indeed, by using properly-built core polarization potentials, instead of explicit consideration of the core electrons, it is possible to reproduce the above-mentioned anomalous features from ab initio calculations in the $A_2$ and $A_2^+$ series.[60–62] The complexity of the used core polarization potentials, and the interplay of various terms within the potential, each with its own stabilizing or destabilizing effect, makes it hard to have a simplified and qualitative picture of the valence-core correlations. One may conclude while the similarity of various positronic bonds to the bonds in lithium dimers and trimers is a promising point to start the bonding analysis, it also implies that the positronic bonds are not regular entities.

Taking this background into account, in this paper we will focus on the two-positron bond formed in $(PsH)_2$, employing ab initio results and analyzing corresponding wavefunctions through various methods. When possible, a comparative analysis is done with $[H^-, e^+, H^-]$ as well as the neutral and cationic lithium dimers. Interestingly, such comparisons reveal the distinct nature of the one- and two-positron bonds and disclose interesting similarities and differences between the positronic bonds and corresponding electronic bonds in the lithium dimers.

## 2. Theory

The MC-QTAIM is an extension of the QTAIM partitioning scheme, proposed originally by Bader and coworkers for the purely electronic molecular systems,[63–71] to the MC quantum systems.[28–32] The MC-QTAIM algorithm partitions a MC quantum system, composed of various types of quantum particles, into the atomic basins, i.e. AIM, in real 3D space using a MC wavefunction, $\Psi_{MC}$. The molecular properties, e.g. molecular expectation values, are also partitioned into the intra-basin, and sometimes into the intra- and inter-basin, contributions. The details of the partitioning algorithm as well as its foundations have been reviewed recently in detail,[34,72] so, only a brief survey is done herein on its main components.

The boundaries between AIM are delineated through the following zero-flux equation: $\vec{\nabla}\Gamma^{(s)}(\vec{r}).\vec{n}(\vec{r}) = 0$, $\Gamma^{(s)}(\vec{r}) = \sum_{n=1}^{s}(m_1/m_n)\rho_n(\vec{r})$, where $\rho_n(\vec{r})$ stands for the one-particle density for $n-th$ type of particles: $\rho_n(\vec{r}) = N_n \int d\tau'_n \Psi^*_{MC}\Psi_{MC}$. $N_n$ is the number of the particles of the $n-th$ type and $d\tau'_n$ implies summing over spin variables of all quantum particles and integrating over spatial coordinates of all quantum particles except one arbitrary particle belonging to the $n-th$ type. $\Gamma^{(s)}$ is the Gamma density while superscript $s$ is called the cardinal number which is the number of distinguishable quantum particles within the quantum system.[30] The Gamma density is the sum of inversely mass-scaled one-particle densities where the masses of the quantum particles are ordered from the lightest, $m_1$, to the heaviest, $m_s$. The solutions of the zero-flux equation are called the zero-flux surfaces where the unit vector, $\vec{n}(\vec{r})$, normal to the surface, is perpendicular to the gradient of the Gamma density, $\vec{\nabla}\Gamma^{(s)}(\vec{r})$, which is tangent to the surface. From all the zero-flux surfaces only those crossing (3, -1) critical points (CPs) act as boundaries between AIM, partitioning the system exhaustively in real 3D space into $P$ atomic basins, $\bigcup_{k}^{P}\Omega_k = R^3$. In the case of the positronic systems the Gamma density is simplified since the masses of electron and positron are equal: $\Gamma^{(2)}(\vec{r}) = \rho_e(\vec{r}) + \rho_p(\vec{r})$ (throughout the text $p$ is used as subscript or superscript to denote position).

Each atomic basin shares a contribution from molecular properties through introducing a property density and integrating the density within the boundaries of the corresponding basin.[34,72] For one-particle properties, represented by the sum of the one-particle operators, $\hat{M} = \sum_{n}^{s}\sum_{i}^{N_n}\hat{m}_{n,i}$, the recipe to build the property density for the $n-th$ type of particles is as follows: $M_n(\vec{r}) = \hat{m}_n \rho_n^{(1)}(\vec{r}',\vec{r})\big|_{\vec{r}=\vec{r}'}$, where $\rho_n^{(1)}(\vec{r}',\vec{r}) = N_n \int d\tau_n' \Psi_{MC}^*(...,\vec{r}_{n,1}',...)\Psi_{MC}(...,\vec{r}_{n,1},...)$ is the generalized form of the reduced first-order density matrix.[31,34,72] The contribution of the $k-th$ basin, $\Omega_k$, to a one-particle property of the $n-th$ type of particles is derived upon integration of the corresponding property density within the boundaries of that basin: $M_n(\Omega_k) = \int_{\Omega_k} d\vec{r}\, M_n(\vec{r})$. The total basin contribution of a one-particle property from all types of particles is: $\tilde{M}(\Omega_k) = \sum_{n}^{s} M_n(\Omega_k)$. The sum of all basin contributions is equal to the expectation value of the total one-particle operator:

$\left\langle \Psi_{MC} \left| \hat{M} \right| \Psi_{MC} \right\rangle = \sum_{k}^{P} \tilde{M}(\Omega_k)$.[34,72] In the case of the two-particle properties, represented by the sum of the two-particle operators, $\hat{G} = \sum_{n}^{s} \left( \sum_{i}^{N_n} \sum_{j>i}^{N_n} \hat{g}_{n,ij} \right) + \sum_{n}^{s} \sum_{m>n}^{s} \left( \sum_{i}^{N_n} \sum_{j}^{N_n} \hat{g}_{nm,ij} \right)$, both the intra-basin, $\tilde{G}(\Omega_k) = \sum_{n}^{s} \left[ G_n(\Omega_k) + \sum_{m>n}^{s} G_{nm}(\Omega_k) \right]$, and inter-basin, $\tilde{G}(\Omega_k, \Omega_l) = \sum_{n}^{s} \left[ G_n(\Omega_k, \Omega_l) + \sum_{m>n}^{s} G_{nm}(\Omega_k, \Omega_l) \right]$, contributions emerge.[72,34] Each term within the brackets is defined as follows: $G_n(\Omega_k) = (1/2) \int_{\Omega_k} d\vec{r}_1 \int_{\Omega_k} d\vec{r}_2 \, \hat{g}_{nn}(\vec{r}_1, \vec{r}_2) \rho_n^{(2)}(\vec{r}_1, \vec{r}_2)$, $G_{nm}(\Omega_k) = \int_{\Omega_k} d\vec{r}_n \int_{\Omega_k} d\vec{r}_m \, \hat{g}_{nm}(\vec{r}_n, \vec{r}_m) \rho_{nm}^{(2)}(\vec{r}_n, \vec{r}_m)$, $G_n(\Omega_k, \Omega_l) = \int_{\Omega_k} d\vec{r}_1 \int_{\Omega_l} d\vec{r}_2 \, \hat{g}_{nn}(\vec{r}_1, \vec{r}_2) \rho_n^{(2)}(\vec{r}_1, \vec{r}_2)$, and $G_{nm}(\Omega_k, \Omega_l) = \int_{\Omega_k} d\vec{r}_n \int_{\Omega_l} d\vec{r}_m \, \hat{g}_{nm}(\vec{r}_n, \vec{r}_m) \rho_{nm}^{(2)}(\vec{r}_n, \vec{r}_m)$. Note that this formalism is developed for two-particle operators that do not include the spatial derivatives which is usually the case. Thus, the formalism only requires to the generalized diagonal form of the reduced second-order density matrix, $\rho_n^{(2)}(\vec{r}_1, \vec{r}_2) = N_n(N_n - 1) \int d\tau_n'' \Psi_{MC}^* \Psi_{MC}$, and the pair density $\rho_{nm}^{(2)}(\vec{r}_n, \vec{r}_m) = N_n N_m \int d\tau_n''' \Psi_{MC}^* \Psi_{MC}$.[31,72,34] Note that $d\tau_n''$ implies summing over spin variables of all quantum particles and integrating over spatial coordinates of all quantum particles except two arbitrary particles, denoted herein as particles "1" and "2", belonging to the $n-th$ type. Also, $d\tau_n'''$ implies summing over spin variables of all quantum particles and integrating over spatial coordinates of all quantum particles except two arbitrary particles, one belonging to the $n-th$ and the other to the $m-th$ types. It is straightforward to demonstrate that the sum of all intra- and inter-basin contributions is equal to the expectation value of the total two-particle operator: $\left\langle \hat{G} \right\rangle = \sum_{k}^{P} \tilde{G}(\Omega_k) + \sum_{k}^{P} \sum_{l>k}^{P} \tilde{G}(\Omega_k, \Omega_l)$.[72,34]

This strategy of partitioning is particularly useful for the total energy partitioning which yields the TC-IQA scheme.

The IQA energy partitioning is a theoretically well-developed and chemically easy to interpret scheme which in recent years has found vast applications in the analysis of AIM interactions in various molecules and molecular aggregates.[38–42] The IQA scheme is the "natural" partitioning of a system wherein the Hamiltonian contains both one- and two-particle operators, e.g. the kinetic energy and the electron-electron interaction energy, respectively. In previous studies, certain terms of the IQA were extended for molecular systems in which a single proton or a positron was treated as a quantum particle on an equal footing with electrons.[35–37] Herein, we will consider the full TC-IQA partitioning scheme conceiving a general TC molecular system where the positronic systems are just one of its specific examples. The general TC system is conceived to be composed of $N$ electrons and $M$ number of another type of quantum particle with the charge $q$ and the mass $m_q$, all within the electric field of $Q$ number of clamped nuclei. A further extension to the MC systems is straightforward and is not considered herein.

The Hamiltonian of the general TC system, written in a.u., is as follows:

$$\hat{H}_{TC} = \hat{T}_e + \hat{T}_q + \hat{V}_{e-nuc} + \hat{V}_{q-nuc} + \hat{V}_{ee} + \hat{V}_{qq} + \hat{V}_{eq} + V_{nuc-nuc}$$

$$= -(1/2)\sum_a^N \nabla_a^2 - (1/2m_q)\sum_c^M \nabla_c^2 - \sum_a^N \sum_A^Q \frac{Z_A}{|\vec{r}_a - \vec{R}_A|} + \sum_c^M \sum_A^Q \frac{qZ_A}{|\vec{r}_c - \vec{R}_A|}$$

$$+ \sum_a^N \sum_{b>a}^N \frac{1}{|\vec{r}_a - \vec{r}_b|} + \sum_c^M \sum_{d>c}^M \frac{q^2}{|\vec{r}_c - \vec{r}_d|} - \sum_a^N \sum_c^M \frac{q}{|\vec{r}_a - \vec{r}_c|} + \sum_A^Q \sum_{B>A}^Q \frac{Z_A Z_B}{|\vec{R}_A - \vec{R}_B|} \quad (1)$$

It is straightforward to demonstrate that the expectation value of the total molecular energy is partitioned as follows:

$$E_{total} = \sum_{i}^{P} E_{intra}(\Omega_i) + \sum_{i}^{P}\sum_{j>i}^{P} E_{inter}(\Omega_i, \Omega_j)$$

$$= \sum_{i}^{P} \left[ T_e(\Omega_i) + T_q(\Omega_i) + V_{e-nuc}(\Omega_i) + V_{q-nuc}(\Omega_i) + V_{ee}(\Omega_i) + V_{qq}(\Omega_i) + V_{eq}(\Omega_i) \right]$$

$$+ \sum_{i}^{P}\sum_{j>i}^{P} \begin{bmatrix} V_{e-nuc}(\Omega_i, \Omega_j) + V_{e-nuc}(\Omega_j, \Omega_i) + V_{q-nuc}(\Omega_i, \Omega_j) + V_{q-nuc}(\Omega_j, \Omega_i) \\ + V_{ee}^{cl}(\Omega_i, \Omega_j) + V_{ee}^{xc}(\Omega_i, \Omega_j) + V_{qq}^{cl}(\Omega_i, \Omega_j) + V_{qq}^{xc}(\Omega_i, \Omega_j) + V_{eq}^{cl}(\Omega_i, \Omega_j) \\ + V_{eq}^{c}(\Omega_i, \Omega_j) + V_{eq}^{cl}(\Omega_j, \Omega_i) + V_{eq}^{c}(\Omega_j, \Omega_i) + V_{nuc-nuc}(\Omega_i, \Omega_j) \end{bmatrix}$$

(2)

The explicit expressions for the intra- and inter-basin terms are as follows:

*Intra − baisn terms*

$$T_e(\Omega_i) = (-1/2) \int_{\Omega_i} d\vec{r}\, \vec{\nabla} \rho_e^{(1)}(\vec{r}',\vec{r})\big|_{\vec{r}'=\vec{r}}, \quad T_q(\Omega_i) = (-1/2m_q) \int_{\Omega_i} d\vec{r}\, \vec{\nabla} \rho_q^{(1)}(\vec{r}',\vec{r})\big|_{\vec{r}'=\vec{r}},$$

$$V_{e-nuc}(\Omega_i) = -Z_i \int_{\Omega_i} d\vec{r}\, \frac{\rho_e(\vec{r})}{|\vec{r}-\vec{R}_i|}, \quad V_{q-nuc}(\Omega_i) = qZ_i \int_{\Omega_i} d\vec{r}\, \frac{\rho_q(\vec{r})}{|\vec{r}_q-\vec{R}_i|},$$

$$V_{ee}(\Omega_i) = (1/2) \int_{\Omega_i} d\vec{r}_1 \int_{\Omega_i} d\vec{r}_2\, \frac{\rho_e^{(2)}(\vec{r}_1,\vec{r}_2)}{|\vec{r}_1-\vec{r}_2|}, \quad V_{qq}(\Omega_i) = (q^2/2) \int_{\Omega_i} d\vec{r}_1 \int_{\Omega_i} d\vec{r}_2\, \frac{\rho_q^{(2)}(\vec{r}_1,\vec{r}_2)}{|\vec{r}_1-\vec{r}_2|},$$

$$V_{eq}(\Omega_i) = -q \int_{\Omega_i} d\vec{r}_e \int_{\Omega_i} d\vec{r}_q\, \frac{\rho_{eq}^{(2)}(\vec{r}_e,\vec{r}_q)}{|\vec{r}_e-\vec{r}_q|},$$

*Inter − baisn terms*

$$V_{e-nuc}(\Omega_i, \Omega_j) = -Z_j \int_{\Omega_i} d\vec{r}\, \frac{\rho_e(\vec{r})}{|\vec{r}-\vec{R}_j|}, \quad V_{e-nuc}(\Omega_j, \Omega_i) = -Z_i \int_{\Omega_j} d\vec{r}\, \frac{\rho_e(\vec{r})}{|\vec{r}-\vec{R}_i|},$$

$$V_{q-nuc}(\Omega_i, \Omega_j) = qZ_j \int_{\Omega_i} d\vec{r}\, \frac{\rho_q(\vec{r})}{|\vec{r}-\vec{R}_j|}, \quad V_{q-nuc}(\Omega_j, \Omega_i) = qZ_i \int_{\Omega_j} d\vec{r}\, \frac{\rho_q(\vec{r})}{|\vec{r}-\vec{R}_i|},$$

$$V_{ee}^{cl}(\Omega_i,\Omega_j) = \int_{\Omega_i} d\vec{r}_1 \int_{\Omega_j} d\vec{r}_2 \frac{\rho_e(\vec{r}_1)\rho_e(\vec{r}_2)}{|\vec{r}_1-\vec{r}_2|}, \quad V_{ee}^{xc}(\Omega_i,\Omega_j) = \int_{\Omega_i} d\vec{r}_1 \int_{\Omega_j} d\vec{r}_2 \frac{\rho_e^{xc}(\vec{r}_1,\vec{r}_2)}{|\vec{r}_1-\vec{r}_2|},$$

$$V_{qq}^{cl}(\Omega_i,\Omega_j) = q^2 \int_{\Omega_i} d\vec{r}_1 \int_{\Omega_j} d\vec{r}_2 \frac{\rho_q(\vec{r}_1)\rho_q(\vec{r}_2)}{|\vec{r}_1-\vec{r}_2|}, \quad V_{qq}^{xc}(\Omega_i,\Omega_j) = q^2 \int_{\Omega_i} d\vec{r}_1 \int_{\Omega_j} d\vec{r}_2 \frac{\rho_q^{xc}(\vec{r}_1,\vec{r}_2)}{|\vec{r}_1-\vec{r}_2|},$$

$$V_{eq}^{cl}(\Omega_i,\Omega_j) = -q \int_{\Omega_i} d\vec{r}_e \int_{\Omega_j} d\vec{r}_q \frac{\rho_e(\vec{r}_e)\rho_q(\vec{r}_q)}{|\vec{r}_e-\vec{r}_q|}, \quad V_{eq}^{c}(\Omega_i,\Omega_j) = -q \int_{\Omega_i} d\vec{r}_e \int_{\Omega_j} d\vec{r}_q \frac{\rho_{eq}^{c}(\vec{r}_e,\vec{r}_q)}{|\vec{r}_e-\vec{r}_q|},$$

$$V_{eq}^{cl}(\Omega_j,\Omega_i) = -q \int_{\Omega_j} d\vec{r}_e \int_{\Omega_i} d\vec{r}_q \frac{\rho_e(\vec{r}_e)\rho_q(\vec{r}_q)}{|\vec{r}_e-\vec{r}_q|}, \quad V_{eq}^{c}(\Omega_j,\Omega_i) = -q \int_{\Omega_j} d\vec{r}_e \int_{\Omega_i} d\vec{r}_q \frac{\rho_{eq}^{c}(\vec{r}_e,\vec{r}_q)}{|\vec{r}_e-\vec{r}_q|},$$

*Assumed mathematical decompositions*

$$\rho_e^{(2)}(\vec{r}_1,\vec{r}_2) = \rho_e(\vec{r}_1)\rho_e(\vec{r}_2) + \rho_e^{xc}(\vec{r}_1,\vec{r}_2),$$

$$\rho_q^{(2)}(\vec{r}_1,\vec{r}_2) = \rho_q(\vec{r}_1)\rho_q(\vec{r}_2) + \rho_q^{xc}(\vec{r}_1,\vec{r}_2),$$

$$\rho_{eq}^{(2)}(\vec{r}_e,\vec{r}_q) = \rho_e(\vec{r}_e)\rho_q(\vec{r}_q) + \rho_{eq}^{c}(\vec{r}_e,\vec{r}_q) \tag{3}$$

In these expressions the superscripts $cl, xc$ and $c$ stand for classical, exchange-correlation and correlation, respectively. The purely electronic intra-basin terms, i.e. $T_e(\Omega_i), V_{e-nuc}(\Omega_i)$ and $V_{ee}(\Omega_i)$, and inter-basin terms, i.e. $V_{e-nuc}(\Omega_i,\Omega_j), V_{e-nuc}(\Omega_j,\Omega_i), V_{ee}^{cl}(\Omega_i,\Omega_j)$ and $V_{ee}^{xc}(\Omega_i,\Omega_j)$, have counterparts in the IQA partitioning scheme. Nevertheless, the terms associated with the second quantum particle, which are delineated by the subscript $q$ (also representing the charge of the particle), i.e. $T_q(\Omega_i)$, $V_{q-nuc}(\Omega_i)$, $V_{qq}(\Omega_i)$, $V_{q-nuc}(\Omega_i,\Omega_j)$, $V_{q-nuc}(\Omega_j,\Omega_i), V_{qq}^{xc}(\Omega_i,\Omega_j)$, are distinctive to the TC-IQA. This is also the case for the terms related to their interactions with electrons, i.e. $V_{eq}(\Omega_i)$, $V_{eq}^{cl}(\Omega_i,\Omega_j)$, $V_{eq}^{c}(\Omega_i,\Omega_j)$, $V_{eq}^{cl}(\Omega_j,\Omega_i)$

, $V_{qq}^{cl}(\Omega_i,\Omega_j)$, $V_{eq}^{c}(\Omega_j,\Omega_i)$. To clarify the role of the second quantum particle in the inter-basin interactions, we categorize the interaction energy contributions into two major classes including the purely electronic class, with counterparts in the IQA, and a new class of terms appearing only due to the presence of the second quantum particle:

$$E_{\text{inter}}(\Omega_i,\Omega_j) = E_{\text{inter}}^{e}(\Omega_i,\Omega_j) + E_{\text{inter}}^{q}(\Omega_i,\Omega_j)$$

$$E_{\text{inter}}^{e}(\Omega_i,\Omega_j) = E_{\text{inter}}^{e,cl}(\Omega_i,\Omega_j) + V_{ee}^{xc}(\Omega_i,\Omega_j)$$

$$E_{\text{inter}}^{e,cl}(\Omega_i,\Omega_j) = V_{e-nuc}(\Omega_i,\Omega_j) + V_{e-nuc}(\Omega_j,\Omega_i) + V_{ee}^{cl}(\Omega_i,\Omega_j) + V_{nuc-nuc}(\Omega_i,\Omega_j)$$

$$E_{\text{inter}}^{q}(\Omega_i,\Omega_j) = E_{\text{inter}}^{q,cl}(\Omega_i,\Omega_j) + V_{qq}^{xc}(\Omega_i,\Omega_j) + E_{\text{inter}}^{eq,c}(\Omega_i,\Omega_j)$$

$$E_{\text{inter}}^{q,cl}(\Omega_i,\Omega_j) = V_{q-nuc}(\Omega_i,\Omega_j) + V_{q-nuc}(\Omega_j,\Omega_i) + V_{qq}^{cl}(\Omega_i,\Omega_j) + V_{eq}^{cl}(\Omega_i,\Omega_j) + V_{eq}^{cl}(\Omega_j,\Omega_i)$$

$$E_{\text{inter}}^{eq,c}(\Omega_i,\Omega_j) = V_{eq}^{c}(\Omega_i,\Omega_j) + V_{eq}^{c}(\Omega_j,\Omega_i) \tag{4}$$

The origins of stabilizing interactions in the IQA scheme are $V_{ee}^{xc}(\Omega_i,\Omega_j)$ and $E_{\text{inter}}^{e,cl}(\Omega_i,\Omega_j)$ terms, usually interpreted as the covalent and the ionic contributions to the bonding, respectively.[43,44] Additionally, in the TC-IQA scheme, depending on the sign of the charge content of the quantum particles new extra "sources" for stabilizing interactions may appear. For positively charged particles, e.g. positrons, these include, $V_{eq}^{cl}(\Omega_j,\Omega_i)$, $V_{qq}^{xc}(\Omega_i,\Omega_j)$ and $E_{\text{inter}}^{eq,c}(\Omega_i,\Omega_j)$, though their relative contributions in each system can only be evaluated numerically. Similar to the case of the purely electronic classical interaction, $E_{\text{inter}}^{q,cl}(\Omega_i,\Omega_j)$ also can be stabilizing (negative) or destabilizing (positive). When the TC-IQA partitioning is done by using a MC-Hartree-Fock (MC-HF) wavefunction, as is done also in this paper, all correlation contributions, e.g. $E_{\text{inter}}^{eq,c}(\Omega_i,\Omega_j)$, disappear and the only possible sources of stabilization may come from

$V_{eq}^{cl}(\Omega_j,\Omega_i)$, $V_{ee}^{x}(\Omega_i,\Omega_j)$ and $V_{qq}^{x}(\Omega_i,\Omega_j)$. In the previous study it was demonstrated that $E_{inter}^{q,cl}(\Omega_i,\Omega_j)$, through the dominance of $V_{eq}^{cl}(\Omega_i,\Omega_j)$, is the origin of the stabilization of the one-positron bond in $[H^-,e^+,H^-]$.[35] Let us also stress that the alternative partitioning of the total energy through applying the extended regional virial theorem,[29,30] while has been applied successfully to the positronic systems,[73] is not particularly informative in regard to the inter-AIM interactions. Accordingly, in this study the TC-IQA energy partitioning is exclusively applied to disclose the nature of inter-AIM interactions.

The MC-QTAIM formalism has been previously applied to certain exotic multi-component systems including the positronic molecules,[35,73–76] and molecules containing positively charged muons.[76–79] It was also applied to systems wherein the proton or its heavier isotopes were treated as quantum particles instead of the clamped point charges.[27,37,76,80] In most of these cases the derived atomic basins, and their computed properties and interaction modes were to a large extent in line with previously given independent arguments regarding their appearance and characteristics.[34,72] In this study we likewise apply the MC-QTAIM and the TC-IQA partitionings to $(PsH)_2$ to shed some light on the nature of the two-positron bond.

### 3. Computational details

#### 3.1. The MC-Hartree-Fock computations

The details of the MC-HF method and the relevant equations have been disclosed elsewhere,[81,82] and in this subsection, only a brief survey is done on their main characteristics for the positronic systems. The MC-HF wavefunction of a positronic system is a product of two Slater determinants, $\Psi_{MC-HF} = \psi_{SD}^{elec.} \cdot \psi_{SD}^{pos.}$, one constructed from the spin-orbitals of electrons, and the other from the spin-orbitals of positrons. For $(PsH)_2$, only singlet spin states were considered for electrons while both singlet and triplet spin states were considered separately for positrons. In the case of the singlet positronic state the unrestricted version of the MC-HF equations were used and two distinct spatial orbitals were attributed to alpha and beta positronic spin states. This yields a spin-contaminated state though the contamination is small around the equilibrium distance. In the case of $[H^-, e^+, H^-]$, the single positron is in a doublet spin state and the electrons were considered only in the singlet spin state. The electronic and positronic spin-orbitals for both species were expanded using the spherical, proton centered, aug-cc-pVTZ basis set, designed for the hydrogen atom,[83] and denoted as [aug-cc-pVTZ:aug-cc-pVTZ]. The MC-HF calculations were redone employing the more extended aug-cc-pVQZ basis set from the same hierarchy but the results did not change significantly (see Table S1 in the supporting information (SI)). Accordingly, the results gained at the MC-HF/[aug-cc-pVTZ:aug-cc-pVTZ] is not far from the complete basis limit and they are used throughout this paper for all subsequent analyzes. $PsH$ was also considered at the same computational levels assuming a singlet electronic spin state. The PySCF program package libraries were used to perform the one- and two-particle integrations, and the self-consistent field procedures in the MC-HF calculations.[84,85]

### 3.2. The Quantum Monte Carlo computations

The full details of the Quantum Monte Carlo (QMC) calculations on $[H^-, e^+, H^-]$ and $(PsH)_2$ species were reported previously and herein the main aspects of its theory and computational implementation are considered.[19,21] The QMC methods can treat the instantaneous correlation between all the particles exactly and on an equal footing, so they are ideal techniques to solve Schrödinger's equation for systems containing positrons. Assuming protons to be clamped for both $[H^-, e^+, H^-]$ and $(PsH)_2$, the following explicitly correlated functional was employed: $\Psi_{QMC} = \hat{A}(1+\hat{i}) \prod_i f_{iA}(r_{iA}) f_{iB}(r_{iB}) \prod_{j>i} g_{ij}(r_{ij})$, as the main building block to describe the two species. In this functional form $f_{iX}(r)$ ($X = A, B$) and $g_{ij}(r)$ are the single-particle and pair correlation functions, respectively, while $\hat{A}$ and $\hat{i}$ are the antisymmetrizer and inversion operators. The variable $r_{iX}$ is the distance between the $i-th$ particle and the clamped proton $X$, and $r_{ij}$ is the distance between $i-th$ and $j-th$ particles. This functional form is the simplest and most general function where all one- and two-body terms have been included.[86,87] The used explicit forms of $f_{iX}(r)$ and $g_{ij}(r)$ are the common generic 3-parameter exponential Padè form: $f_{iX}(r) = g_{ij}(r) = Exp\left[(ar + br^2)/(1 + cr)\right]$. The parameters, $a, b$ and $c$, are determined independently for each function and may assume different values during the energy optimization procedure. To reduce the number of the nonlinear variational parameters and to help their optimization, parameter $a$ was fixed according to the cusp condition, to theoretical values of +1 and –1 for the one-particle positron-proton and electron-proton functions, respectively.[86,87] For the electron-electron and the positron-positron pair correlation functions $a = 1/2$ and $a = 1/4$ were used for unlike and like spin particles, respectively, while for the electron-positron pair correlation function $a = -1/2$ was fixed.[88] For both species the electrons were assumed to be in a singlet spin state with zero total angular momentum. The positrons were assumed to be in a singlet spin state for $(PsH)_2$ whereas the single positron in $[H^-, e^+, H^-]$ was considered in a doublet spin state. The wave functions, for different values of the inter-proton distances, have been optimized using variational Monte Carlo (VMC) as was also done previously.[19,21] In the previous studies, using the fixed-node Diffusion Monte Carlo (DMC) the potential energy curves (PECs) were computed in 0.4 Bohr steps and the equilibrium distances were located at 6.4 ± 0.4 Bohr (the total energy is -1.3403 Hartree) for $[H^-, e^+, H^-]$ and 6.0 ± 0.4 Bohr (the total energy is -1.5888 Hartree) for $(PsH)_2$.[19,21] The DMC energies were estimated using 3000 walkers and a time step of 0.003 Hartree$^{-1}$. All QMC calculations suffer from statistical noise, so each point on the PECs has an error bar and for this reason, it is difficult to accurately pinpoint the position of the minimum of the PEC. Furthermore, the wave functions of each point of the PEC are independently optimized using a stochastic algorithm, adding further difficulties to the goal of obtaining smooth PECs. To partially minimize this effect and to gain better accuracy, in the present study, the PECs were scanned in finer 0.1 Bohr steps from 5.9 to 6.9 for $[H^-, e^+, H^-]$ and from 5.6 to 6.4 Bohr for $(PsH)_2$. The used fixed nodal surfaces of the wavefunction were always those optimized at 6.4 and 6.0 Bohr for $[H^-, e^+, H^-]$ and $(PsH)_2$, respectively.

### 3.3. The radial pair distribution functions of the DMC wavefunction

The spatial structure of $(PsH)_2$ was considered employing the radial pair distribution functions (RPDF) of various pair of particles, derived from the DMC wavefunction at the equilibrium distance, i.e. at 6.0 Bohr. The effectiveness of these functions to elucidate the details of the structure of exotic positronic species has been demonstrated previously.[88–90] Since the basic formalism of the RPDF has been discussed comprehensively previously,[91,92] herein, only a brief survey is done on its basic traits. A PDF is introduced between any pair of particles as follows: $\rho_{ij}(\vec{r}) = \int \delta(\vec{r}_{ij} - \vec{r}) \Psi^* \Psi d\tau$, wherein $d\tau$ implies integration on all the spin and spatial variables within the wavefunction and $\vec{r}_{ij}$ is the position vector between $(i, j)$ pair. One of the particles in the pair could be a clamped point charge, e.g. proton in $(PsH)_2$ or a body-fixed point in space, e.g. center of mass. In such cases the PDF is practically a one-particle distribution relative to a fixed point and is called after the name of the corresponding quantum particle, e.g. the electronic PDF. In spherically symmetric states one may further simplify the formulation and replace the position vector with the radial variable in the polar coordinates.[92] Since the two protons in $(PsH)_2$ are treated as clamped point charges, this simplification is not applicable herein, and the PDFs are orientation-dependent thus, the axis going through the two protons is employed as a natural choice for the identification of fixed reference points. It is customary for graphical representation to depict RPDFs,[88–90,92] $D_{ij}(\vec{r}) = Nr^2 \rho_{ij}(\vec{r})$, instead of PDFs, where $N$ is the normalization factor. Since there are four electrons and two positrons in $(PsH)_2$, the electronic and positronic RPDFs have been normalized to 4 and 2, respectively. For comparison purposes, the RPDFs of $Ps_2$ and $PsH$ were also computed using both VMC and DMC employing 3000 walkers and, for DMC, a time step of 0.003 Hartree$^{-1}$. Since $Ps_2$ is composed of two electrons and two positrons, the corresponding RPDFs have been normalized to 2 and according to the same reasoning the electronic and positronic

RPDFs of $PsH$ have been normalized to 2 and 1, respectively. At the next stage of the analysis, the total electronic RPDFs of $PsH$ and $(PsH)_2$ were partitioned into "separate-electron" and "separate-positron" distributions. This type of partitioning was originally proposed by Cox and coworkers for the purely electronic systems,[92] and since then has been extended by Bressanini to the positronic systems as well.[89,90] The partitioning is done by considering that, for each electronic configuration in real space, the electrons may be ordered according to their distance from a fixed point, e.g. a clamped nucleus. For instance, in the case of $PsH$ there is one electron closer to the nucleus (the "inner" electron) and one more distant from the nucleus (the "outer" electron). The RPDF of the inner electron is defined as follows: $D_{ij}^{in}(\vec{r}) = Nr^2 \int \delta(\min(\vec{r}_{12}, \vec{r}_{13}) - \vec{r}) \Psi^* \Psi d\tau$, in which the indices 2 and 3 refer to electrons and 1 to the clamped proton. The RPDF of the outer electron is $D_{ij}^{out}(\vec{r}) = Nr^2 \int \delta(\max(\vec{r}_{12}, \vec{r}_{13}) - \vec{r}) \Psi^* \Psi d\tau$ and it is easy to demonstrate that $D_{ij}(\vec{r}) = D_{ij}^{in}(\vec{r}) + D_{ij}^{out}(\vec{r})$.[90,92] It is straightforward to extend this approach to multi-electron/positron species and to partition the RPDFs into separate-particle contributions. Let us stress that this partitioning scheme does not violate the principle of indistinguishability of electrons since electrons are not labeled themselves. For instance, in each specific configuration in real space, electrons 1 or 2 may exchange their role as the inner or outer electron of $PsH$ according to their relative distance to the proton. Eventually, this partitioning scheme was applied to $(PsH)_2$ and the results were compared to those derived for $PsH$.

### 3.4. The MC-QTAIM and TC-IQA analysis of the MC-HF wavefunction

The topological analysis was done on the Gamma density of $(PsH)_2$, derived from the MC-HF/[aug-cc-pVTZ:aug-cc-pVTZ] wavefunction at the MC-HF and DMC derived equilibrium distances, 8.4 and 6.0 Bohr, respectively, and the AIM structure was derived. The electronic and positronic populations of the $k-th$ basin were deduced by direct integration of the corresponding one-particle densities: $N_e(\Omega_k) = \int_{\Omega_k} \rho_e(\vec{r}) d\vec{r}$, $N_p(\Omega_k) = \int_{\Omega_k} \rho_p(\vec{r}) d\vec{r}$ ($k=1,2$). The basin polarization electronic and positronic dipole vectors were also deduced as follows: $\vec{P}_e(\Omega_k) = -\int_{\Omega_k} \vec{r}_k \rho_e(\vec{r}_k) d\vec{r}_k$, $\vec{P}_p(\Omega_k) = \int_{\Omega_k} \vec{r}_k \rho_p(\vec{r}_k) d\vec{r}_k$, wherein $\vec{r}_k$ is the distance of electrons/positrons relative to the proton retained in the $k-th$ basin.[29] The energy components of each basin and the nature of the inter-basin interactions were derived at both equilibrium distances using the TC-IQA method by direct integration as detailed in the previous section. Part of the MC-QTAIM and the TC-IQA computations and the artwork of Figure 6 were done using the AIMAll package, T. A. Keith, AIMAll (Version 13.11.04), TK Gristmill Software, Overland Park KS, USA, 2013.

## 4. Results and discussion

### 4.1. Ab initio data

The PEC of $(PsH)_2$ was scanned from 3.2 to 12.0 Bohr in 0.4 Bohr steps at the MC-HF level and then, to locate the equilibrium distance more precisely, a fine grid scan was done from 8.0 to 8.8 Bohr in 0.1 Bohr steps (see Table S1 in the SI). The equilibrium distance was found to be at 8.4 ± 0.1 Bohr for the singlet positronic spin state (the total energy is -1.3314 Hartree) though the PEC is quite flat around this point. Using the same strategy and at the same computational level the equilibrium distance of $\left[H^-,e^+,H^-\right]$ was located at 6.9 ± 0.1 Bohr (the total energy is -1.1741 Hartree) (see Table S2 in the SI). This is fairly similar to what was reported previously by Reyes and coworkers namely an equilibrium distance of 7.0 Bohr (the total energy -1.1755 Hartree) albeit derived using a more extended positronic basis set.[17] Our attempts to locate an equilibrium distance for the triplet positronic spin states of $(PsH)_2$ failed. The resulting PEC was found to be monotonically decaying by increasing the distance between the two protons, similar to the previously gained results with highly accurate ab initio calculations.[21] This is the first indication that the positrons, not electrons, are responsible for the bonding between two $PsH$ atoms in $(PsH)_2$.

At the DMC level, the equilibrium distances were located at 6.3 ± 0.1 Bohr (the total energy is -1.3405 ± 0.0001 Hartree) for $\left[H^-,e^+,H^-\right]$ and 6.0 ± 0.1 Bohr for the singlet positronic spin state of $(PsH)_2$ (the total energy is -1.5887 ± 0.0001 Hartree) while all the energies are guaranteed to be an upper bound of the exact energies (see Tables S1 and S2 in the SI). Like the case of the PECs derived at the MC-HF level, here too, the PECs are relatively flat around the equilibrium distances. To see whether the PECs may contain any bound vibrational level, the one-dimensional nuclear Schrödinger equation was solved employing the Numerov method for the previously mentioned PECs, derived at 0.4 Bohr steps,[19,21] using the mathematical procedure disclosed recently.[37] All vibrational energy levels, below the energy of the proper dissociation limit, i.e. $(PsH)_2 \to 2PsH$ and $\left[H^-,e^+,H^-\right] \to PsH + H^-$, were derived. The PEC of $(PsH)_2$ contains only six vibrational levels whereas that of $[H^-, e^+, H^-]$ contains at least ten levels but the zero-point energies of both species are virtually equal, ~2.9 kJ.mol$^{-1}$ (see Table S3 in the SI for the numerical values).

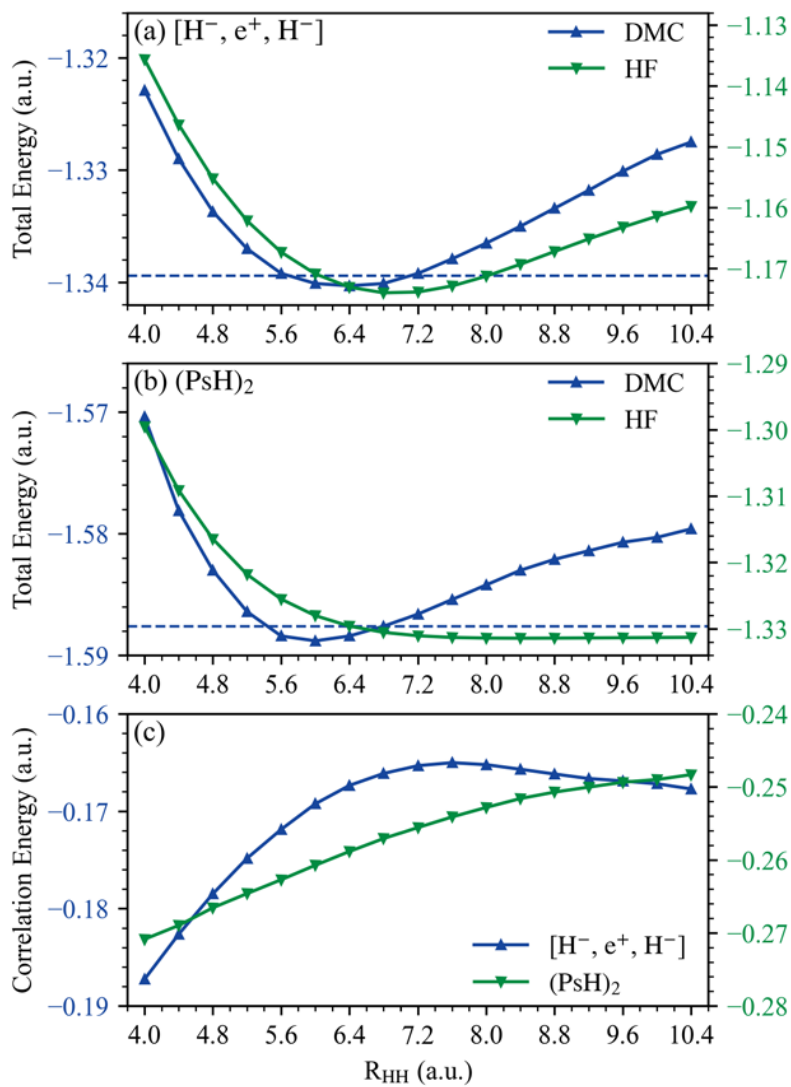

Fig 1. Ground-state PECs of (a) $[H^-, e^+, H^-]$ and (b) $(PsH)_2$ computed at the DMC and MC-HF/[aug-cc-pVTZ:aug-cc-pVTZ] levels. The dashed lines in panels (a) and (b) indicate the first vibrational energy level computed for the DMC derived PECs. (c) The correlation energies computed as the difference between the DMC and MC-HF energies.

At the first step of the analysis in order to understand the role of correlations in $[H^-, e^+, H^-]$ and $(PsH)_2$, panels (a) and (b) of Figure 1 offers the corresponding PECs computed at the DMC (hereafter assumed to be practically exact) and the MC-HF levels. For $(PsH)_2$, it is evident that the two PECs are quite distinct and the predicted equilibrium distances are 2.4 Bohr apart. At the MC-HF level the minimum is located at a larger distance compared to the exact equilibrium distance and the PEC is very shallow and goes correctly toward the dissociation limit of two independent/non-interacting $PsH$ atoms. The BDEs computed at the corresponding equilibrium distances, $E_{BDE}(R_{eq.}) = E_{(PsH)_2}(R_{eq.}) - 2 \times E_{PsH}$, are ~ -27 and ~ -1 kJ.mol$^{-1}$ at the DMC and MC-HF levels, respectively. Evidently, the inclusion of the correlation energy is of great importance for the stabilization of $(PsH)_2$ relative to the separate $PsH$ atoms. The absolute difference between the two computed PECs at the DMC and MC-HF levels is the total absolute correlation energy, $E_{corr.}(R_{HH}) = |E_{exact}(R_{HH}) - E_{MC-HF}(R_{HH})|$, which has been depicted in panel (c) of Figure 1. The total correlation energy of $(PsH)_2$ in the whole depicted range of the inter-proton distances is relatively constant and varies between ~ 650 and ~ 710 kJ.mol$^{-1}$. At the same range of distances, the total correlation energy of $[H^-, e^+, H^-]$ varies between ~ 430 and ~ 490 kJ.mol$^{-1}$ and it is appreciably smaller than that of $(PsH)_2$. From a percentage perspective, these amount to ~ 30% to ~ 35% increase in total correlation energy upon the addition of a single positron to $[H^-, e^+, H^-]$ and the formation of $(PsH)_2$. It must be stressed that in contrast to purely electronic systems, there are three types of correlations in the two-component quantum systems.[93–97] In the case of the positronic systems, these include the electron-electron, the electron-positron and the positron-positron correlations. In the present study we will not attempt to disentangle the contribution of each of these correlations in the total correlation since as discussed elsewhere,[98,99] this is quite a non-trivial computational task. However, the above-mentioned percentage increase of the correlation energy point to non-negligible contribution of the positron-related correlations in $(PsH)_2$.

One may also compare the change of the total correlation energy in the process of forming $(PsH)_2$ from infinitely distant $PsH$ atoms. Based on our computed total DMC and MC-HF energies, the total correlation energy of $PsH$ atom (See Table S4 in the SI), and $(PsH)_2$ at 8.4 Bohr, are ~ 325 and ~ 661 kJ.mol$^{-1}$, respectively. Thus, the net change of correlation energy for the process of $(PsH)_2$ formation is ~ 11 kJ.mol$^{-1}$ which only amounts to ~ 2% of the sum of the correlation energies of two $PsH$ atoms. This small percentage is an indirect hint that $PsH$ atoms are only mildly perturbed within $(PsH)_2$ and retain their original structure to a large extent. This is an intriguing result since the total correlation energy generally depends on the number of correlated particles and the positron-positron correlation interaction only appears in $(PsH)_2$. Let us now compare $(PsH)_2$ with some other relevant species.

In the case of $[H^-, e^+, H^-]$, as is also evident from Figure 1, the equilibrium distance computed at the MC-HF level is ~ 0.6 Bohr longer than the exact result.

However, the PEC at the MC-HF level approach the proper dissociation limit of the independent $PsH$ and $H^-$ species. The BDEs computed at the corresponding equilibrium distances are ~ -62 and ~ -55 kJ.mol$^{-1}$ at the DMC and MC-HF levels, respectively. Evidently, the contribution of the correlation energy in BDE is not as substantial as it is in the case of $(PsH)_2$. Let us now compare $(PsH)_2$ to $Li_2$ as its candidate dual.

The exact high-level equilibrium inter-nuclear distance and BDE of $Li_2$ are ~ 5.05 Bohr and ~ -102 kJ.mol$^{-1}$, while those computed at the HF/aug-cc-pVTZ level are ~ 5.26 Bohr and ~ -17 kJ.mol$^{-1}$.[100,101] The total correlation energies of $Li$ atom, and $Li_2$ at 5.05 Bohr, are ~ 119 and ~ 325 kJ.mol$^{-1}$, respectively. The net change in the correlation energy in the process of $Li_2$ formation is ~ 87 kJ.mol$^{-1}$, which is ~ 37% of the sum of the correlation energies of the two $Li$ atoms. This percentage and the computed BDEs are both significantly larger in $Li_2$ relative to $(PsH)_2$ while the equilibrium distance of $Li_2$ is smaller than that of $(PsH)_2$. All these support the view that the two-electron bond in $Li_2$ is considerably stronger than the two-positron bond in $(PsH)_2$.

From a classical viewpoint $Li_2$ is a combination of two electronic cores, each with a unit of positive charge, and two valence electrons, while $(PsH)_2$ is composed of two electronic cores, each with a unit of negative charge, and two valence positrons. The two cores have very different polarizabilities and ionization potentials, ~ 34 Å$^3$ and ~ 73 kJ.mol$^{-1}$ for $H^-$ versus ~ 0.03 Å$^3$ and ~ 7298 kJ.mol$^{-1}$ for $Li^+$.[50,102,103] Evidently, the electrons composing the $Li^+$ are strongly bonded and form a compact core with an effective radius ~ 0.9 Å, whereas those composing $H^-$ are loosely bonded and form a rather diffuse core within an effective radius, ~ 1.3 Å.[104] In contrast to these differences, in a sense, there is a kind of duality between the two species. However, quantum mechanically, electrons are indistinguishable fermions obeying Pauli's exclusion principle. This imposes the constraint of the orthogonalization of the valence sigma molecular orbital of $Li_2$, composed mainly of the linear combination of the lithium 2s atomic orbitals, to the two core orbitals, composed mainly of the lithium 1s atomic orbitals,[52] yielding a nodal structure in the valence orbital.[59] The core-valence orbital orthogonalization and associated exchange interactions must be always taken into account in any realistic analysis of the Alkali dimers,[53–59,105–107] or constructing any reliable core polarization potential.[60–62] Because of the distinguishability of electrons and positrons, this constraint and the relevant exchange interactions are absent in the case of the positronic orbital of $(PsH)_2$. It is not a trivial task to disentangle the quantitative effects of the orthogonalization and exchange interactions on the BDE and the equilibrium distance of $Li_2$. By the way, it seems safe to claim that this discussion points to a "qualitative" distinction between $(PsH)_2$ and $Li_2$ making the above-mentioned duality only partly justifiable in a quantum mechanical analysis. These observations do not inevitably dismiss similarities between bonds in $Li_2$ and $(PsH)_2$ but point to the distinctive character of the two-positron bond in the latter.

The same argument is also applicable in comparison of $[H^-,e^+,H^-]$ and $Li_2^+$ as species with one-particle bonds. The exact equilibrium inter-nuclear distance and the BDE of $Li_2^+$ are ~5.87 Bohr and ~ -125 kJ.mol$^{-1}$, while those computed at the HF/aug-cc-pVTZ level are ~5.92 Bohr and ~ -123 kJ.mol$^{-1}$.[108,109] The computed BDE of $Li_2^+$ is twice larger than that of $\left[H^-,e^+,H^-\right]$'s BDE while the equilibrium distance is appreciably longer in $\left[H^-,e^+,H^-\right]$. The effect of the correlation energy is negligible on the equilibrium distances and the BDEs of both species. These results both point to the similarities and distinctiveness of the one-electron and one-positron bonds in these two species.

Finally, let us stress that like the case of $Li_2/Li_2^+$ pair, $(PsH)_2/\left[H^-,e^+,H^-\right]$ pair also reveals the same anomalous bonding features, namely, both exact values of the BDE and equilibrium distance in $\left[H^-,e^+,H^-\right]$ are larger than the corresponding ones in $(PsH)_2$ i.e. stronger but longer bond in $\left[H^-,e^+,H^-\right]$ compared to $(PsH)_2$. In their succinct but thoughtful paper on this anomaly,[53] Buckingham and Rowlands noted that upon the addition of an electron to the valence sigma molecular orbital of $Li_2^+$, and the formation of $Li_2$, the energies of both valence and core molecular orbitals increase significantly. This was interpreted as an indication of large and comparable destabilizing core-valence and valence-valence interactions that make the addition of the second valence electron energetically unfavorable. The increase of the molecular orbital energies in $Li_2^+ + e^- \rightarrow Li_2$ process, computed at their respective equilibrium distances at (RO)HF/aug-cc-pVTZ levels, are ~ 583, ~ 583 and ~ 611 kJ.mol$^{-1}$, for the two core and the valence orbitals, respectively. In the case of the analogous process in the positronic duals, $\left[H^-,e^+,H^-\right] + e^+ \rightarrow (PsH)_2$, the energies of the core orbitals "decrease" by ~ 449 and ~ 501 kJ.mol$^{-1}$ whereas the energy of the positronic orbital "increases" by ~ 437

kJ.mol$^{-1}$. Evidently, the "destabilizing" core-valence interactions in $Li_2$ change into "stabilizing" core-valence interactions in $(PsH)_2$, while the valence-valence interactions are destabilizing in both species. This is understandable since the added positron, in contrast to the electron, has a positive charge and its interaction with the core electrons is expected to be stabilizing. Indeed, in a similar process upon the addition of a positron to the hydride ion, $H^- + e^+ \rightarrow PsH$, the energy of the only involved electronic orbital also "decreases" significantly, ~ 655 kJ.mol$^{-1}$. The comparison of the above-mentioned processes seemingly indicates that if the core-valence interactions are important in $Li_2$, they must be also important in the case of $(PsH)_2$. However, the trends of the changes of the core-valence and valence-valence interactions upon the addition of a positron to $[H^-, e^+, H^-]$ are conflicting. These conflicting trends make it hard to apply the argument proposed by Buckingham and Rowlands, at least in its original form, to explain the anomalous bonding of $(PsH)_2 / [H^-, e^+, H^-]$ pair and this is compatible with the fact that the anomalous trend disappears at the MC-HF level. We leave a more thorough analysis of the positron-positron and positron-core electron interactions in $(PsH)_2$ to the last subsection. We conclude this subsection once again emphasizing that while the supposed duality of $Li_2 / Li_2^+$ and $(PsH)_2 / [H^-, e^+, H^-]$ pairs is conceivable as a proper initial guide, it must not mask the unique characteristics of the positronic bonds.

### 4.2. Deriving the spatial structure of $(PsH)_2$ from the DMC wavefunction

During the analysis of the variation of the total correlation energy upon the formation of $(PsH)_2$ from $PsH$ atoms, it was tentatively concluded that the latter ones are mildly perturbed within $(PsH)_2$. Theoretically, two limits of weakly and strongly interacting $PsH$ atoms may be conceived where only the former option justifies the notation $(PsH)_2$. In the strong interaction limit, as will be discussed, the two $Ps$ atoms may be combined to form a dipositronium,[110,111] $Ps_2$, and $(PsH)_2$ is better represented as $[H, Ps_2, H]$. In this subsection, to test these options independently from the previous analysis and to have a more direct view into details of the internal spatial structure of $(PsH)_2$, various RPDFs computed from the DMC wavefunction at the equilibrium distance, i.e. 6 Bohr, are considered. Panel (a) in Figure 2 depicts the electronic and positronic RPDFs along the axis going through the two protons and relative to the center located in the middle of the two protons. The maximum of the electronic RPDF is located at ~ 3 Bohr, which is the distance of the protons from the origin, confirming that the electrons are mainly localized around the protons although with a large width of distribution. In the case of the positronic RPDF the maximum is located further at ~ 4 Bohr, beyond the clamped proton, and the distribution is even wider than that of the electronic RPDF. It is also noticeable from this panel that between 0 to 1 Bohr the electronic and positronic RPDFs coalesce. This is a hint that between the two protons, a positron is always accompanied by an electron, which is consistent with the usual qualitative picture of $PsH$ as a hydrogen atom with a $Ps$ "orbiting" around.[45–48] This picture is also supported upon comparing the one-electron and one-positron densities of $PsH$, depicted in Figure 3, as will be discussed in the next subsection.

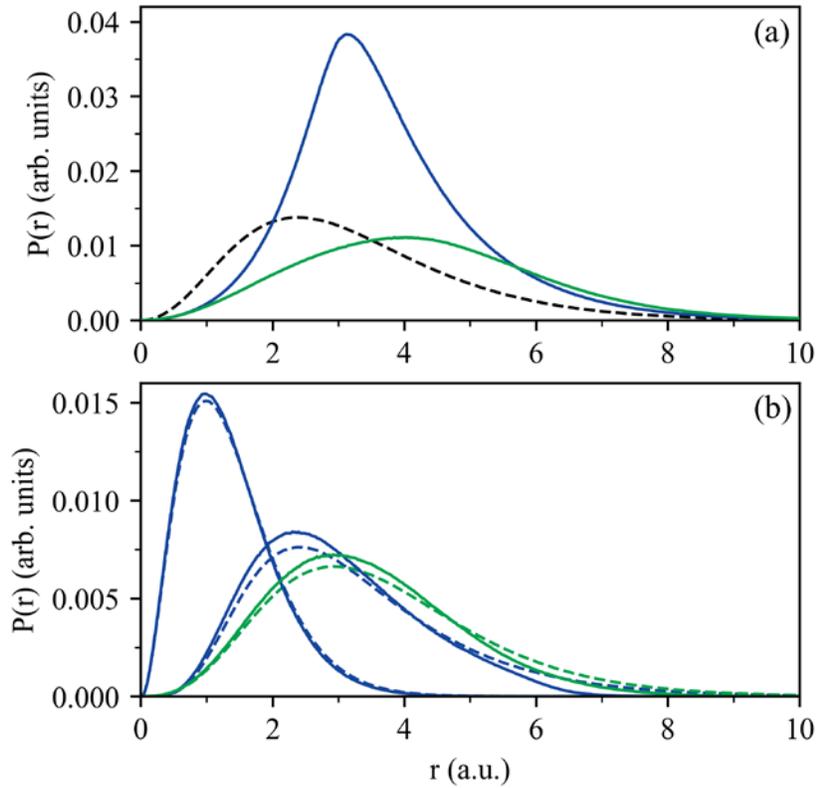

Fig 2. (a) Electronic (blue) and positronic (green) radial pair distribution functions depicted with respect to the middle of the inter-proton axis in $(PsH)_2$ (the clamped proton is located at 3 Bohr) compared to the same distributions in $Ps_2$, which are depicted with respect to its center of mass (dashed black). (b) The separate-electron (blue) and separate-positron (green) radial pair distribution functions of $PsH$ (dashed curves) and $(PsH)_2$ (solid curves) with respect to a proton in each species (the proton used as the reference in $(PsH)_2$ is that placed in the left-hand side of the middle of the inter-proton axis). Only the two closest electrons and the closest positron are plotted for $(PsH)_2$.

In analogy with the bond formation between two hydrogen atoms where the two electrons are shared forming an electron pair, an intriguing hypothesis is that in $(PsH)_2$ the two $Ps$ atoms are shared by the two $PsH$ atoms, forming $Ps_2$.[110,111] It is conceivable that the partial formation of a $Ps_2$ "substructure" between the two protons could enhance the stability of the system, forming $[H, Ps_2, H]$. To investigate this hypothesis the same panel also offers the electronic and positronic RPDFs of $Ps_2$ with respect to the center of mass. Because of the high symmetry of $Ps_2$, the two RPDFs are superimposable and have been depicted as a single dashed curve. Evidently, both electronic and positronic RPDFs of $(PsH)_2$ are rather different than the corresponding ones in $Ps_2$. In the case of the validity of the hypothesis of $Ps_2$ formation between the two protons, one should expect that at least the positronic RPDFs to be similar in $(PsH)_2$ and $Ps_2$. However, they are clearly different with maxima located at different positions and one may conclude that there is no sign of the formation of a $Ps_2$ substructure within $(PsH)_2$. Now let us proceed to verify the hypothesis that in $(PsH)_2$ there are two marginally perturbed $PsH$ atoms.

Panel (b) in Figure 2 depicts, in dashed lines, the separate-electron and the positronic RPDFs of $PsH$ depicted assuming the clamped proton as the center of the coordinate system. The inner electron distribution strictly resembles the hydrogen atom's electron distribution while the outer electron is closer, on average, to the positron. In order to compare the separate-particle RPDFs in $(PsH)_2$ with those of $PsH$, the RPDFs were collected assuming the proton placed in the left-hand side of the middle point of the inter-proton axis as the reference point. Accordingly, for each configuration, during the QMC simulations, the distances of all particles were ordered from the closest to the farthest with respect to the leftmost proton. This means the electronic/positronic RPDFs were partitioned into the sum of four/two separate distributions although only the RPDFs of the two electrons and the one positron closest to the proton are depicted as solid lines in Figure 2(b). Inspection of the figure demonstrates that the RPDF of the closest electron to the proton in $(PsH)_2$ is almost superimposable to the RPDF of the inner electron in $PsH$. The next two separate-electron and separate-positron RPDFs are also very similar to the corresponding ones in $PsH$ and their maxima are almost in the same positions. However, in a more detailed view these RPDFs are a little more localized in $(PsH)_2$ than $PsH$. This means that the probability to find the positron and the accompanying electron between the two protons is a little higher than the corresponding one in $PsH$. Since the equilibrium distance of $(PsH)_2$ is ~ 6 Bohr, while the maximum of the separate-positron RPDF in $PsH$ is ~ 3 Bohr, the positron tends to stay naturally in the middle of the protons. All these suggest a picture of $(PsH)_2$ as a union of two distorted $PsH$ atoms where the inner electron of each proton is virtually unperturbed, while the remaining electron-positron pair of each $PsH$ are shared but without forming a $Ps_2$. By the way, to check the nature of this sharing and see whether it may help to stabilize $(PsH)_2$, through forming a bond between two $PsH$ atoms, further analysis is needed which is done in the next subsection. We conclude this subsection once again emphasizing that $(PsH)_2$ is formed from two mildly perturbed $PsH$ atoms and in the next subsection we will try to shed some light on the nature of interactions between the two $PsH$ atoms.

### 4.3. The MC-QTAIM analysis of the MC-HF wavefunction

Before considering the results of the MC-QTAIM partitioning of $(PsH)_2$ it is worth discussing the status of the $PsH$ atom within the context of the MC-QTAIM as well as the variation of the one-electron and one-positron densities upon the formation of $(PsH)_2$ from $PsH$ atoms. Figure 3 offers the one-electron, the one-positron and the Gamma densities of $PsH$ at the MC-HF/[aug-cc-pVTZ:aug-cc-pVTZ] level. The Gamma density has a single maximum at the proton, i.e. a single (3, -3) CP, which is the typical topological structure of a single atom,[67] justifying the term $PsH$ "atom". Figure 4 offers the one-electron and one-positron difference maps of $(PsH)_2$ with respect to two non-interacting $PsH$ atoms at the same computational level as well as the corresponding difference maps for $H_2$ and $Li_2$ molecules compared to their respective atomic constituents. A one-electron difference map is defined as the difference between the one-electron density of a molecule at a certain geometry and the sum of the ground state one-electron densities of the constituent non-interacting atoms at the same geometry.[112] Mathematically, the difference map is introduced as follows: $\Delta\rho_e(\vec{r}) = \rho_e^{mol.}(\vec{r}) - \sum_i^P \rho_e^{atom,i}(\vec{r})$, and the same may easily be extended to the one-positron density difference maps: $\Delta\rho_p(\vec{r}) = \rho_p^{mol.}(\vec{r}) - \sum_i^P \rho_p^{atom,i}(\vec{r})$. The idea of using the one-electron density difference maps as a tool to trace the bonding interactions between AIM has a long and sometimes controversial history in both computational and experimental communities.[113–123] Comparison of panels (a) and (b) of Figure 4 clearly reveals the pattern of $\Delta\rho_e(\vec{r})$ is quite different in $H_2$ and $Li_2$ in line with previous reports.[114,117,124] In the case of $H_2$ the difference map reveals a simple pattern of the one-electron density accumulation between the protons and its depletion outside this region. On the other hand, in the case $Li_2$ in addition to the one-electron density accumulation between the nuclei, two new regions of density accumulation also appear behind the lithium nuclei.

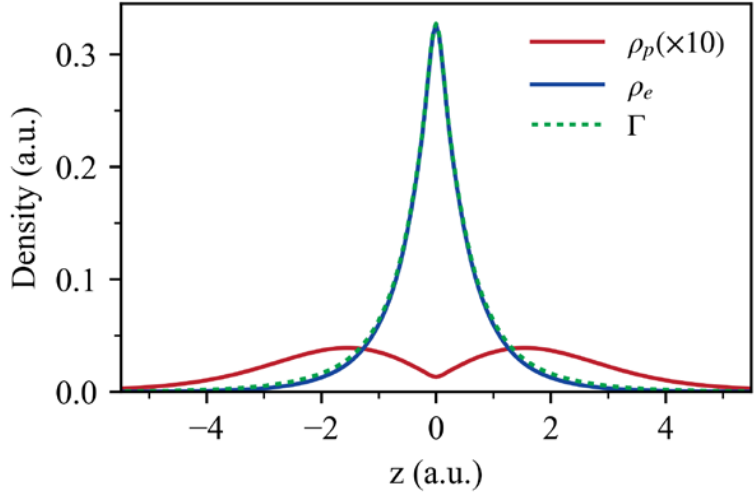

Fig 3. 1D depictions of the Gamma (dashed green), the one-electron (blue) and the one-positron (red) densities of $PsH$ atom. The clamped proton is at the center of the coordinate system. The one-positron density is scaled ten times larger for better visibility.

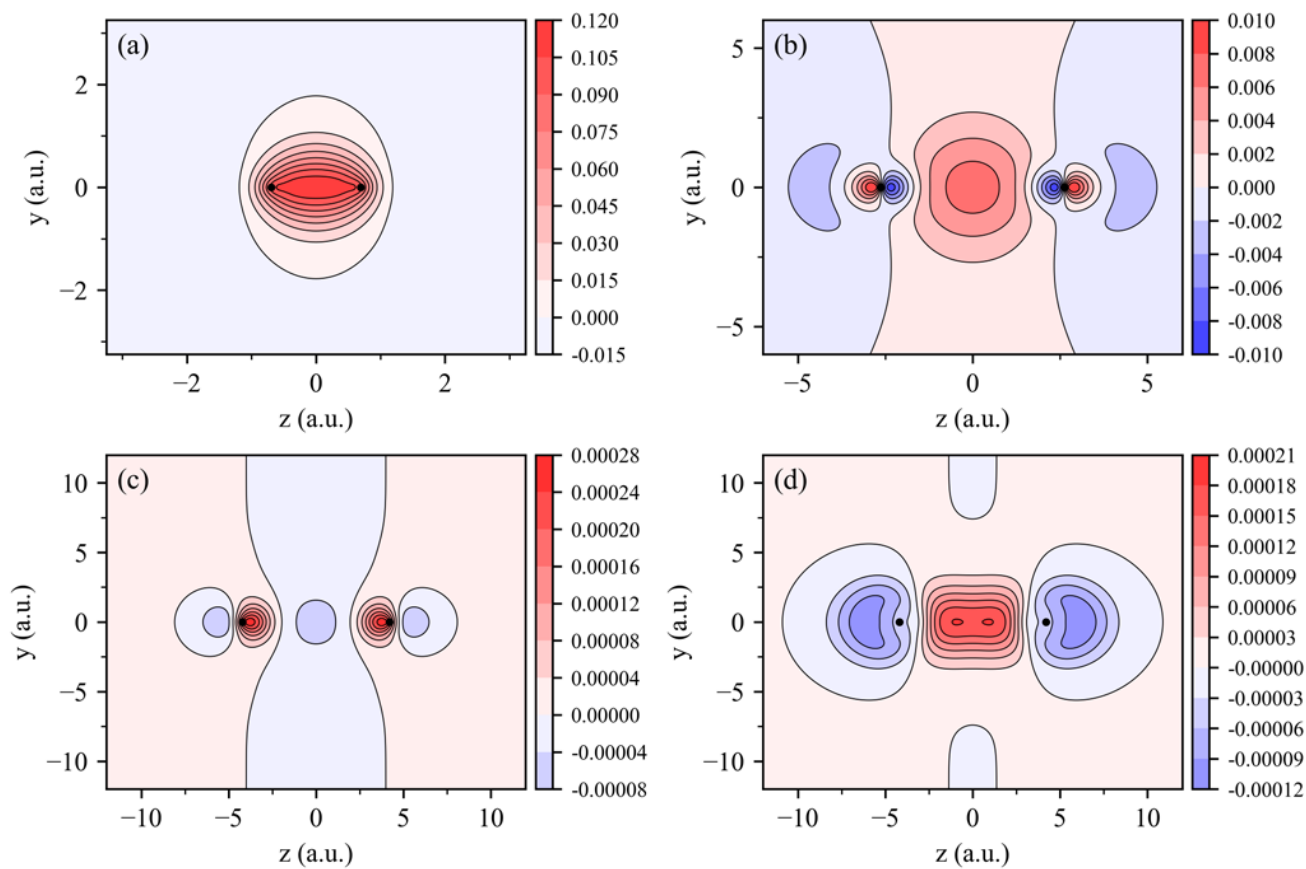

Fig 4. Difference maps of (a) the one-electron density in $H_2$, (b) the one-electron density of $Li_2$, (c) the one-electron density of $(PsH)_2$ and (d) the one-positron density of $(PsH)_2$. The positions of the clamped proton in each system are designated by black dots. All maps have been computed at the corresponding molecular equilibrium geometries. The equilibrium inter-nuclear distances for $H_2$ and $Li_2$ are 1.387 and 5.261 Bohr at the RHF/aug-cc-pVTZ level of theory, respectively. The one-electron densities of hydrogen and lithium atoms were also computed at the (U)HF/aug-cc-pVTZ level for consistency.

In between these two accumulation regions, there are also two regions of the density depletion in front of each nucleus, revealing the polarization of the two cores of $Li_2$. Schwarz and coworkers have attributed the more convoluted pattern of the density depletion and accumulation in $Li_2$ and the core polarization to the previously discussed orthogonalization constraint of the valence to the core orbitals.[117] They interpreted this pattern to be the result of "non-bonded repulsion between inner shell and the bonding orbitals" in $Li_2$.[117] Panel (c) in Figure 4 depicts $\Delta\rho_e(\vec{r})$ for $(PsH)_2$ which reveals that the pattern of core polarization is more similar to that observed for $Li_2$ rather than $H_2$ once again confirming that the former is a better dual of $(PsH)_2$ though, the distinctive features of $(PsH)_2$ are also clearly observable. For example, the one-electron density at the midpoint between the two protons of $(PsH)_2$ at the equilibrium distance, given in a.u., is ~ 0.0010 while the sum of the one-electron densities of two non-interacting $PsH$ atoms at the same point is minutely larger ~ 0.0011. The difference is less than 0.0001 and this means more than 95% of the one-electron density is recoverable from the one-electron densities of non-interacting $PsH$ atoms and no charge accumulation is observable at the midpoint between the two protons. In comparison, the one-electron density at the midpoint between the two nuclei of $Li_2$ is ~ 0.013 and it is more than 50% larger than the sum of the one-electron densities of two $Li$ atoms, ~ 0.006, at the same point in space. Also, if one compares panels (b) and (c), it is evident that the pattern of the core polarization is "reversed" in $(PsH)_2$ in comparison with $Li_2$. This reversed polarization is in line with the one-positron density map depicted in panel (d), which demonstrates the one-positron density of $(PsH)_2$ accumulates in between, and depleted behind, the two protons. Evidently, the cores are polarized in a way that electron accumulation is closer to the positron accumulation region between the two protons, benefiting from the stabilizing interaction between particles carrying opposite charges. This is another manifestation of the previously mentioned change of the core-valence destabilizing interactions in $Li_2$ into stabilizing interactions in $(PsH)_2$. Let us now discuss the results of the MC-QTAIM and the TC-IQA partitionings performed using the MC-HF/[aug-cc-pVTZ:aug-cc-pVTZ] wavefunction at the DMC and MC-HF equilibrium distances, 6.0 and 8.4 Bohr, respectively.

Panels (a) and (b) in Figure 5 depict the relief maps of the one-positron density and the corresponding Laplacian of the one-positron density that acts like a magnifying glass, revealing the regions of concentration and depletion of the one-positron density at 8.4 Bohr inter-proton distance (See Figure S1 for the same results at 6.0 Bohr).[67] The patterns reveal the diffuse concentration of positrons between protons in line with panel (d) of Figure 4 as well as the avoidance of positrons around the protons. Similar pattern was previously observed in the analysis of $\left[H^-, e^+, H^-\right]$,[35] while as is evident from Figure S1, at 6.0 Bohr the concentration is even more pronounced between the protons. Panels (c) and (d) in Figures 5 and S1 depict the relief maps of the one-electron and the Gamma densities revealing their profound similarities. This is understandable since the one-positron density is much more diffuse than the one-electron density and contributes marginally to the Gamma density as the sum of the one-positron and the one-electron densities. The topography of the Gamma density is typical of a diatomic species, with two (3, -3) CPs and a (3, -1) CP in between, while the flat zero-flux surface of the gradient of the Gamma density goes through (3, -1) CP and divides the system evenly into two AIM. Figure 6 depicts the shape of the two equivalent atomic basins each containing half of the populations of electrons, $N_e(\Omega_1) = N_e(\Omega_2) = 2$, and positrons, $N_p(\Omega_1) = N_p(\Omega_2) = 1$.

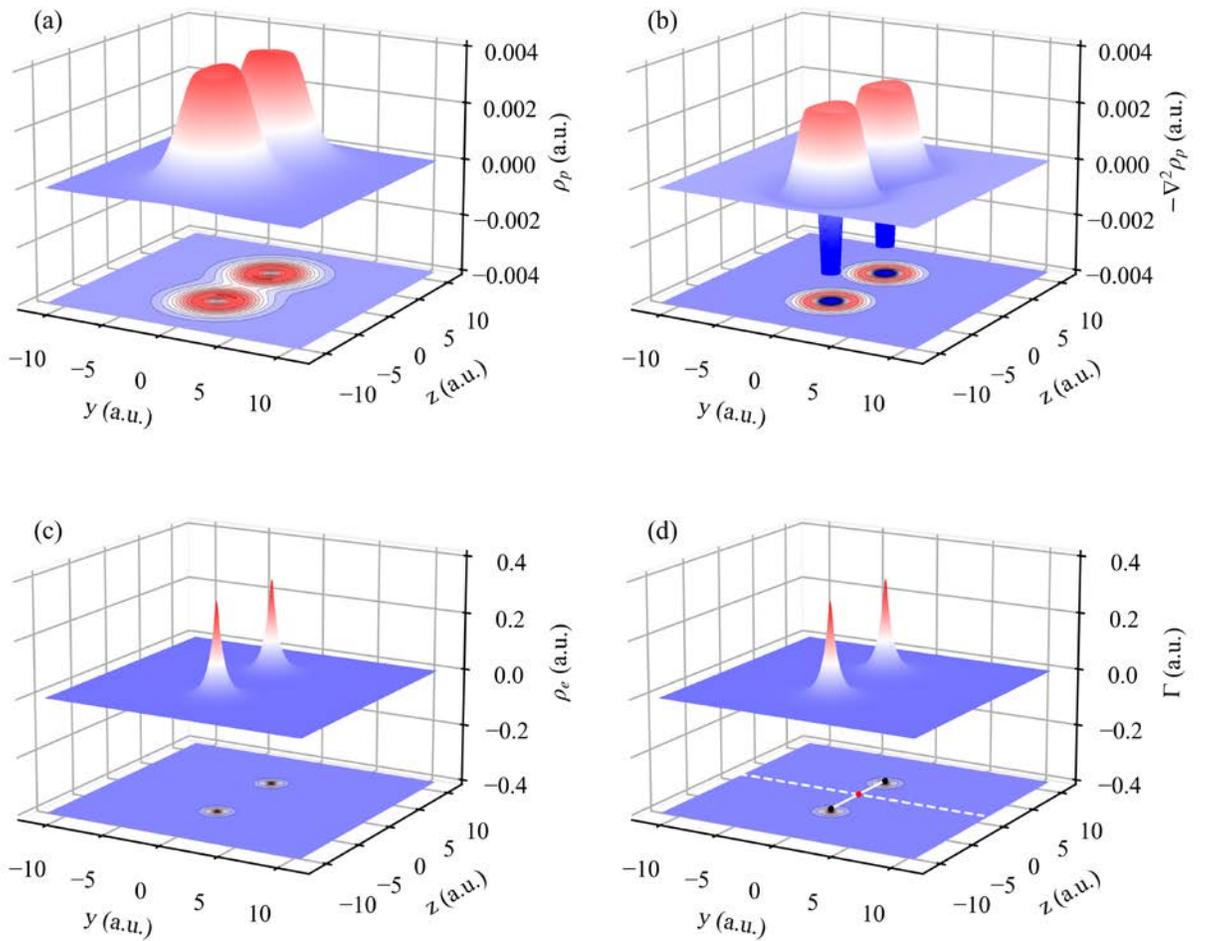

Fig 5. Relief maps of (a) the one-positron density, (b) the Laplacian of the one-positron density, (c) the one-electron density, and (d) the Gamma density of $(PsH)_2$. The protons are 8.4 Bohr apart, placed at (0.0, 0.0, 4.2) and (0.0, 0.0, -4.2) in the coordinate system. The black and red spheres in panel (d) are the (3, -3) and (3, -1) CPs of the Gamma density, respectively, while the solid white lines are the gradient paths connecting the (3, -3) and (3, -1) CPs. The dashed white line is the intersection of the zero-flux surface and the yz plane, which acts as the boundary of the AIM.

There is no charge transfer between the two atomic basins and each basin has on average exactly the same number of the electrons and positrons of an isolated $PsH$ atom; thus, these basins may be viewed as "deformed" $PsH$ atoms. Let us now consider the properties and interaction modes of these two AIM.

To have a quantitative view of the extent of deformation of these basins, the basin polarization electronic and positronic dipole vectors were computed at 8.4 Bohr for each basin as the measures of basin deformation. The magnitudes of the electronic vectors, for both basins, all given in a.u., are equal, $\left|\vec{P}_e(\Omega_1)\right| = \left|\vec{P}_e(\Omega_2)\right| \sim 0.03$, and smaller than the magnitude of the positronic vectors, $\left|\vec{P}_p(\Omega_1)\right| = \left|\vec{P}_p(\Omega_2)\right| \sim 0.08$. All these vectors are null in the isolated "not deformed" $PsH$ atom because of its isotropic one-particle densities depicted in Figure 3. The dominant component of both electronic and positronic vectors lies on the axis connecting the two protons and the contributions from the two other perpendicular components are practically null. The directions of the electronic dipoles point away from the middle point between the two protons, whereas the situation is reversed in the case of the positronic dipoles. The direction and magnitude of the electronic dipoles are both in accord with the intra-basin core polarization depicted in Figure 4(c). The accumulation and depletion regions of the positronic density, depicted in Figures 4(d) and 5(a,b), are farther from the nuclei compared to the corresponding regions for the electronic counterparts. This is also in line with the direction and the magnitude of the computed positronic dipoles. It would be instructive to compare all these results with the deformation induced in $Li$ basins of $Li_2$ at its equilibrium distance, as the formal dual of $(PsH)_2$. Alas, the AIM structure of $Li_2$ is not similar to the typical two-basin structure of most diatomic species and it is composed of three atomic basins instead of two.[125–128] The one-electron density of $Li_2$ possesses an extra (3, -3) CP in the middle of the two $Li$ nuclei, i.e. a non-nuclear maximum which forms an atomic basin without containing any nucleus. This basin is usually called a "pseudo-atom" and shares boundaries with the two $Li$ basins so there are two (3, -1) CPs each between one of the $Li$ basins and the pseudo-atom.[129–131] This is another example of a distinctive feature between $(PsH)_2$ and $Li_2$ in contrast to the proposed duality.

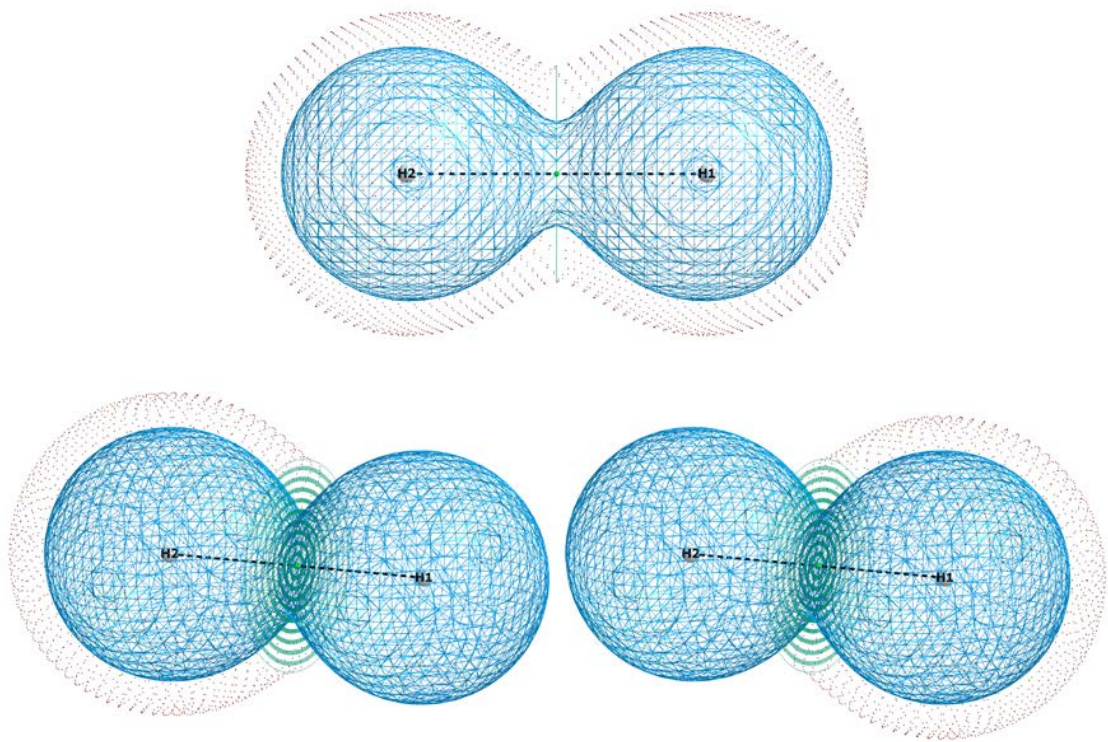

Fig 6. AIM of $(PsH)_2$ depicted for 8 Bohr inter-proton distance. The blue spherical mesh is the 3D iso-density surface of the one-positron density shown at 0.0015 a.u. while the violet mesh indicates surfaces where Gamma density is equal to 0.001 a.u., which are arbitrary but reasonable outer boundaries for the atomic basins. The flat inter-atomic surface that acts as the boundary between AIM are shown by dark green surfaces. Small light green spheres are the (3, -1) CPs and the black dashed lines are the gradient paths connecting (3, -1) and (3, -3) CPs. The latter CPs are located at or very near to the protons which are shown by the larger gray spheres.

In order to quantify the interaction between the two atomic basins in $(PsH)_2$, the TC-IQA analysis was done at 6.0 and 8.4 Bohr inter-proton distances (see Tables S5 and S6 in the SI for the individual values of the TC-IQA components). The values of the two main interaction classes, given in a.u., are as follows: $E^e_{inter}(\Omega_1,\Omega_2) = 0.1146$ and $E^p_{inter}(\Omega_1,\Omega_2) = -0.1224$ at 8.4 Bohr, demonstrating that the source of stabilization is in the positronic interaction energy term. To unravel the role of these interactions on the previously reported BDE, -1.0 kJ.mol$^{-1}$, it was partitioned according to eq. (2) as follows: $E_{BDE} = E_{def.}(\Omega_1) + E_{def.}(\Omega_2) + E_{inter}(\Omega_1,\Omega_2)$. The basin "deformation" energy, $E_{def.}(\Omega_i) = E_{intra}(\Omega_i) - E_{PsH}$, is the energetic manifestation of the reorganization induced within one atomic basin upon the interaction with the other basin.[39,40] The basin deformation energies are equal, $E_{def.}(\Omega_1) = E_{def.}(\Omega_2)$, and their sum is: ~ +19.6 kJ.mol$^{-1}$, which for a larger extent compensates for the total inter-basin interaction energy, ~ -20.5 kJ.mol$^{-1}$ yielding a small BDE. Interestingly, even at 6 Bohr also the total interaction term is stabilizing, ~ -91.4 kJ.mol$^{-1}$, but the sum of the two deformation energies, ~ +99.3 kJ.mol$^{-1}$, outweighs the former and their sum is ~ +8.0 kJ.mol$^{-1}$ larger than the energy of two independent *PsH* atoms. What is the mechanism behind the stabilizing inter-basin interaction? To answer this question let us consider all the interaction terms at 8.4 Bohr, given in a.u., in detail: $E^{e,cl}_{inter}(\Omega_1,\Omega_2) = 0.1174$, $V^x_{ee}(\Omega_1,\Omega_2) = -0.0028$, $E^{p,cl}_{inter}(\Omega_1,\Omega_2) = -0.1172$, $V^x_{pp}(\Omega_1,\Omega_2) = -0.0052$. The value of the electronic exchange term, $V^x_{ee}(\Omega_1,\Omega_2)$, ~ -7.4 kJ.mol$^{-1}$, which is usually perceived as a tracer of the covalent bonds,[43] is smaller than that computed previously for $[H^-, e^+, H^-]$, ~ -38 kJ.mol$^{-1}$.[35]

The evident smallness of this term when compared to those computed for the covalently bonded molecules rules out the presence of any appreciable electronic covalent character for the bond in $(PsH)_2$.[43] The large destabilizing classical electronic interactions, $E_{inter}^{e,cl}(\Omega_i,\Omega_j)$, ~ 308.4 kJ.mol$^{-1}$, rules out also any ionic character to the bond. We conclude that purely electronic terms cannot be responsible for bonding between $PsH$ atoms of $(PsH)_2$ and the bond is mainly a positronic bond. Thus, the mechanism behind the formation of the positronic bond must be traced in the two remaining terms responsible for stabilizing interactions, i.e. $V_{pp}^x(\Omega_1,\Omega_2)$ and $E_{inter}^{p,cl}(\Omega_1,\Omega_2)$. The value of the positronic exchange term, $V_{pp}^x(\Omega_1,\Omega_2)$, ~ -13.6 kJ.mol$^{-1}$, is almost two times larger than its electronic counterpart but its absolute value is not large compared to the typical values of $V_{ee}^x(\Omega_1,\Omega_2)$ in the covalently bonded molecules.[43] Since the value of $E_{inter}^{p,cl}(\Omega_1,\Omega_2)$, ~ -307.8 kJ.mol$^{-1}$, is much larger than that of $V_{pp}^x(\Omega_1,\Omega_2)$, the electrostatic character of the positronic bond seems to be the dominant contribution in the positronic bond of $(PsH)_2$. The origin of the electrostatic character is the classical stabilizing interaction between the one-positron density, encompassed in one of the basins, with the one-electron density, encompassed in the other basin. The sum of the corresponding terms, $V_{ep}^{cl}(\Omega_1,\Omega_2)+V_{ep}^{cl}(\Omega_2,\Omega_1)$, ~ -1212.5 kJ.mol$^{-1}$, dominates $E_{inter}^{p,cl}(\Omega_1,\Omega_2)$, and is the major driving force behind the formation of the position bond. This was also the same electrostatic mechanism of bonding disclosed in the case of formation of the one-positron bond in $[H^-,e^+,H^-]$.[35]

We conclude that the positronic bond in $(PsH)_2$ has a mainly novel electrostatic character and because of the distinctiveness between these electrostatic interactions and the usual ionic interactions, which are operative in both the considered one- and two-positron bonds, these bonds deserve to be categorized as a new class of bonds. This class is particularly different from the purely covalent two-positron bonds formed, for example, in anti-hydrogen molecule made of anti-hydrogen atom,[132,133] in which the two positrons hold the two anti-protons near each other similar to what the electrons do to protons in a normal hydrogen molecule. To emphasize the novel character of the one- and two-positron bonds in $[H^-, e^+, H^-]$ and $(PsH)_2$, we propose to call it the "gluonic" character. This terminology seems to be appropriate since in this class of bonds positrons act as a "classical" glue and bond two otherwise repelling hydrides. Whether the gluonic character is something unique, as seen in a handful of exotic positronic systems, or more widespread, is an interesting question that needs to be answered in future studies.

## 5. Conclusion and prospects

The analysis done in this paper reveals that the considered positronic bonds belong to a novel category of bonds where the mechanism of bonding is electrostatic in nature but quite distinct from the usual ionic bonds. This new type of electrostatic interaction is foreign to purely electronic bonds and has only been traced in the above-mentioned one- and two-positron bonds and a recently disclosed exotic bond in the quantum superposed malonaldehyde.[37] Thus, not only the study of positronic bonds extends the concept of bonding beyond the boundaries of electronic matter, but also reveals a novel mechanism of bonding. While all these are promising, we believe these are just the first steps to a completely new and uncharted territory. The following topics are particularly of interest for future studies.

The three-positron bond is the next natural target for a comprehensive bonding analysis.[22] It seems reasonable to assume that its dual in the lithium diatomic series is $Li_2^-$ anion and an MC-QTAIM analysis may help to uncover the nature of bonding in this species. The three-center two-positron bonds are also worth considering in detail since as emphasized by Reyes and coworkers,[49] these are linked to fundamental chemical concepts like $\sigma$- aromaticity/anti-aromaticity and pseudo-metallic character. In fact, the unequivocal discovery of any form of anti-matter aromaticity may extend the concept beyond purely electronic matter, which in itself is an intriguing possibility. At the technical level, extending the MC-QTAIM analysis to correlated wavefunctions is an inevitable step to gain more concrete results about the nature of bonding in the positronic species. Particularly interesting is to uncover the energetic contributions of each type of the three mentioned correlations to the positronic bonds. All these and related problems are now under study in our labs and the results will be disclosed in future reports.

# References

[1] L. Pauling, *The Nature of the Chemical Bond and the Structure of Molecules and Crystals: An Introduction to Modern Structural Chemistry*, Cornell University Press, Ithaca, New York, **1960**.
[2] C. A. Russell, *The History of Valency*, Leicester University Press, Leicester, **1971**.
[3] R. S. Mulliken, *Chem. Rev.* **1931**, *9*, 347–388.
[4] J. H. Van Vleck, A. Sherman, *Rev. Mod. Phys.* **1935**, *7*, 167–228.
[5] K. Ruedenberg, *Rev. Mod. Phys.* **1962**, *34*, 326–376.
[6] W. Kutzelnigg, *Angew. Chem. Int. Ed. Engl.* **1973**, *12*, 546–562.
[7] C. A. Coulson, *Coulson's Valence*, Oxford University Press, Oxford ; New York, **1979**.
[8] T. A. Albright, J. K. Burdett, M.-H. Whangbo, *Orbital Interactions in Chemistry*, Wiley-Interscience, Hoboken, New Jersey, **2013**.
[9] G. Frenking, S. Shaik, Eds. , *The Chemical Bond: Fundamental Aspects of Chemical Bonding*, Wiley-VCH, **2014**.
[10] G. Frenking, S. Shaik, Eds. , *The Chemical Bond: Chemical Bonding Across the Periodic Table*, Wiley-VCH, Weinheim, **2014**.
[11] L. Zhao, M. Hermann, W. H. E. Schwarz, G. Frenking, *Nat. Rev. Chem.* **2019**, *3*, 48–63.
[12] G. N. Lewis, *J. Am. Chem. Soc.* **1916**, *38*, 762–785.
[13] I. Langmuir, *J. Am. Chem. Soc.* **1919**, *41*, 868–934.
[14] G. N. Lewis, *Valence and the Structure of Atoms and Molecules*, The Chemical Catalog Company, **1923**.
[15] L. Zhao, W. H. E. Schwarz, G. Frenking, *Nat. Rev. Chem.* **2019**, *3*, 35–47.
[16] D. Griffiths, *Introduction to Elementary Particles*, Wiley-VCH, Weinheim, **2008**.
[17] J. Charry, M. T. do N. Varella, A. Reyes, *Angew. Chem. Int. Ed.* **2018**, *57*, 8859–8864.
[18] S. Ito, D. Yoshida, Y. Kita, M. Tachikawa, *J. Chem. Phys.* **2020**, *153*, 224305.
[19] D. Bressanini, *J. Chem. Phys.* **2021**, *154*, 224306.
[20] F. Moncada, L. Pedraza-González, J. Charry, M. T. do N. Varella, A. Reyes, *Chem. Sci.* **2019**, *11*, 44–52.
[21] D. Bressanini, *J. Chem. Phys.* **2021**, *155*, 054306.
[22] D. Bressanini, *J. Chem. Phys.* **2022**, *156*, 154302.
[23] G. F. Gribakin, J. A. Young, C. M. Surko, *Rev. Mod. Phys.* **2010**, *82*, 2557–2607.
[24] K. Süvegh, T. Marek, in *Handb. Nucl. Chem.* (Eds.: A. Vértes, S. Nagy, Z. Klencsár, R.G. Lovas, F. Rösch), Springer US, Boston, MA, **2011**, pp. 1461–1484.
[25] F. Tuomisto, I. Makkonen, *Rev. Mod. Phys.* **2013**, *85*, 1583–1631.
[26] J. Mitroy, M. W. J. Bromley, G. G. Ryzhikh, *J. Phys. B At. Mol. Opt. Phys.* **2002**, *35*, R81–R116.
[27] M. Goli, S. Shahbazian, *Theor. Chem. Acc.* **2011**, *129*, 235–245.
[28] M. Goli, S. Shahbazian, *Theor. Chem. Acc.* **2012**, *131*, 1208.
[29] M. Goli, S. Shahbazian, *Theor. Chem. Acc.* **2013**, *132*, 1362.
[30] M. Goli, S. Shahbazian, *Theor. Chem. Acc.* **2013**, *132*, 1365.
[31] M. Goli, S. Shahbazian, *Theor. Chem. Acc.* **2013**, *132*, 1410.
[32] M. Gharabaghi, S. Shahbazian, *J. Chem. Phys.* **2017**, *146*, 154106.

[33] S. Shahbazian, in *Adv. Quantum Chem. Topol. QTAIM* (Eds.: J.I. Rodríguez, F. Cortés-Guzmán, J.S.M. Anderson), Elsevier, **2023**, pp. 73–109.
[34] S. Shahbazian, *Compr. Comput. Chem. Ser. Ref. Module Chem. Mol. Sci. Chem. Eng.* **2022**, 1–13.
[35] M. Goli, S. Shahbazian, *ChemPhysChem* **2019**, *20*, 831–837.
[36] M. Goli, S. Shahbazian, *ChemPhysChem* **2016**, *17*, 3875–3880.
[37] M. Goli, S. Shahbazian, *Phys. Chem. Chem. Phys.* **2023**, *25*, 5718–5730.
[38] P. L. A. Popelier, D. S. Kosov, *J. Chem. Phys.* **2001**, *114*, 6539–6547.
[39] M. A. Blanco, A. Martín Pendás, E. Francisco, *J. Chem. Theory Comput.* **2005**, *1*, 1096–1109.
[40] E. Francisco, A. Martín Pendás, M. A. Blanco, *J. Chem. Theory Comput.* **2006**, *2*, 90–102.
[41] A. F. Silva, L. J. Duarte, P. L. A. Popelier, *Struct. Chem.* **2020**, *31*, 507–519.
[42] J. M. Guevara-Vela, E. Francisco, T. Rocha-Rinza, Á. Martín Pendás, *Molecules* **2020**, *25*, 4028.
[43] M. García-Revilla, E. Francisco, P. L. A. Popelier, A. M. Pendás, *ChemPhysChem* **2013**, *14*, 1211–1218.
[44] A. M. Pendás, J. L. Casals-Sainz, E. Francisco, *Chem. – Eur. J.* **2019**, *25*, 309–314.
[45] J. Mitroy, *Phys. Rev. Lett.* **2005**, *94*, 033402.
[46] F. Rolim, T. Moreira, J. R. Mohallem, *Braz. J. Phys.* **2004**, *34*, 1197–1203.
[47] S. L. Saito, *Nucl. Instrum. Methods Phys. Res. Sect. B Beam Interact. Mater. At.* **2000**, *171*, 60–66.
[48] D. Bressanini, G. Morosi, *J. Chem. Phys.* **2003**, *119*, 7037–7042.
[49] J. Charry, F. Moncada, M. Barborini, L. Pedraza-González, M. T. do N. Varella, A. Tkatchenko, A. Reyes, *Chem. Sci.* **2022**, *13*, 13795–13802.
[50] A. S. McNeill, K. A. Peterson, D. A. Dixon, *J. Chem. Phys.* **2020**, *153*, 174304.
[51] A. Miffre, M. Jacquey, M. Büchner, G. Trénec, J. Vigué, *Phys. Rev. A* **2006**, *73*, 011603.
[52] A. Haaland, *Molecules and Models: The Molecular Structures of Main Group Element Compounds*, Oxford University Press, Oxford ; New York, **2008**.
[53] A. D. Buckingham, T. W. Rowlands, *J. Chem. Educ.* **1991**, *68*, 282.
[54] A. C. Roach, P. Baybutt, *Chem. Phys. Lett.* **1970**, *7*, 7–10.
[55] F. Koike, H. Nakamura, *Chem. Phys. Lett.* **1978**, *53*, 31–34.
[56] M. Pelissier, E. R. Davidson, *Int. J. Quantum Chem.* **1984**, *25*, 723–731.
[57] E. Lombardi, L. Jansen, *Phys. Rev. A* **1986**, *33*, 2907–2912.
[58] G. Dotelli, E. Lombardi, L. Jansen, *J. Mol. Struct. THEOCHEM* **1992**, *276*, 159–166.
[59] F. Weinhold, C. R. Landis, *Valency and Bonding: A Natural Bond Orbital Donor-Acceptor Perspective*, Cambridge University Press, Cambridge, UK ; New York, **2005**.
[60] W. Müller, J. Flesch, W. Meyer, *J. Chem. Phys.* **1984**, *80*, 3297–3310.
[61] W. Müller, W. Meyer, *J. Chem. Phys.* **1984**, *80*, 3311–3320.
[62] I. Schmidt-Mink, W. Müller, W. Meyer, *Chem. Phys.* **1985**, *92*, 263–285.
[63] R. F. W. Bader, T. T. Nguyen-Dang, *Adv. Quantum Chem.* **1981**, *14*, 63–124.
[64] R. F. W. Bader, T. T. Nguyen-Dang, Y. Tal, *Rep. Prog. Phys.* **1981**, *44*, 893–948.
[65] R. F. W. Bader, *Acc. Chem. Res.* **1985**, *18*, 9–15.

[66] R. F. W. Bader, *Chem. Rev.* **1991**, *91*, 893–928.
[67] R. F. W. Bader, *Atoms in Molecules: A Quantum Theory*, Clarendon Press, **1994**.
[68] P. L. Popelier, *Atoms in Molecules: An Introduction*, Prentice Hall, Harlow, **2000**.
[69] C. F. Matta, R. J. Boyd, *The Quantum Theory of Atoms in Molecules: From Solid State to DNA and Drug Design*, Wiley-VCH, Weinheim, **2007**.
[70] J. I. Rodriguez, F. Cortés-Guzmán, J. S. M. Anderson, Eds. , *Advances in Quantum Chemical Topology Beyond QTAIM*, Elsevier, **2022**.
[71] Á. M. Pendás, E. Francisco, D. Suárez, A. Costales, N. Díaz, J. Munárriz, T. Rocha-Rinza, J. M. Guevara-Vela, *Phys. Chem. Chem. Phys.* **2023**, *25*, 10231–10262.
[72] S. Shahbazian, in *Adv. Quantum Chem. Topol. QTAIM* (Eds.: J.I. Rodriguez, F. Cortés-Guzmán, J.S.M. Anderson), Elsevier, **2022**, pp. 73–109.
[73] M. Goli, S. Shahbazian, *Int. J. Quantum Chem.* **2011**, *111*, 1982–1998.
[74] P. Nasertayoob, M. Goli, S. Shahbazian, *Int. J. Quantum Chem.* **2011**, *111*, 1970–1981.
[75] F. Heidar Zadeh, S. Shahbazian, *Int. J. Quantum Chem.* **2011**, *111*, 1999–2013.
[76] M. Goli, S. Shahbazian, *Phys. Chem. Chem. Phys.* **2015**, *17*, 245–255.
[77] M. Goli, S. Shahbazian, *Phys. Chem. Chem. Phys.* **2014**, *16*, 6602–6613.
[78] M. Goli, S. Shahbazian, *Phys. Chem. Chem. Phys.* **2015**, *17*, 7023–7037.
[79] M. Goli, S. Shahbazian, *Chem. - Eur. J.* **2016**, *22*, 2525–2531.
[80] M. Goli, S. Shahbazian, *Comput. Theor. Chem.* **2015**, *1053*, 96–105.
[81] M. Tachikawa, K. Mori, H. Nakai, K. Iguchi, *Chem. Phys. Lett.* **1998**, *290*, 437–442.
[82] S. P. Webb, T. Iordanov, S. Hammes-Schiffer, *J. Chem. Phys.* **2002**, *117*, 4106–4118.
[83] T. H. Dunning, *J. Chem. Phys.* **1989**, *90*, 1007–1023.
[84] Q. Sun, T. C. Berkelbach, N. S. Blunt, G. H. Booth, S. Guo, Z. Li, J. Liu, J. D. McClain, E. R. Sayfutyarova, S. Sharma, S. Wouters, G. K.-L. Chan, *WIREs Comput. Mol. Sci.* **2018**, *8*, e1340.
[85] Q. Sun, X. Zhang, S. Banerjee, P. Bao, M. Barbry, N. S. Blunt, N. A. Bogdanov, G. H. Booth, J. Chen, Z.-H. Cui, J. J. Eriksen, Y. Gao, S. Guo, J. Hermann, M. R. Hermes, K. Koh, P. Koval, S. Lehtola, Z. Li, J. Liu, N. Mardirossian, J. D. McClain, M. Motta, B. Mussard, H. Q. Pham, A. Pulkin, W. Purwanto, P. J. Robinson, E. Ronca, E. R. Sayfutyarova, M. Scheurer, H. F. Schurkus, J. E. T. Smith, C. Sun, S.-N. Sun, S. Upadhyay, L. K. Wagner, X. Wang, A. White, J. D. Whitfield, M. J. Williamson, S. Wouters, J. Yang, J. M. Yu, T. Zhu, T. C. Berkelbach, S. Sharma, A. Yu. Sokolov, G. K.-L. Chan, *J. Chem. Phys.* **2020**, *153*, 024109.
[86] D. Bressanini, G. Morosi, *J. Chem. Phys.* **2003**, *119*, 7037–7042.
[87] D. Bressanini, G. Morosi, *J. Phys. B At. Mol. Opt. Phys.* **2008**, *41*, 145001.
[88] D. Bressanini, *Phys. Rev. A* **2019**, *99*, 022510.
[89] D. Bressanini, *Phys. Rev. A* **2018**, *97*, 012508.
[90] D. Bressanini, *Phys. Rev. A* **2021**, *104*, 022819.
[91] E. Mátyus, J. Hutter, U. Müller-Herold, M. Reiher, *J. Chem. Phys.* **2011**, *135*, 204302.
[92] A. W. King, L. C. Rhodes, H. Cox, *Phys. Rev. A* **2016**, *93*, 022509.
[93] H. Nakai, *Int. J. Quantum Chem.* **2007**, *107*, 2849–2869.

[94] T. Ishimoto, M. Tachikawa, U. Nagashima, *Int. J. Quantum Chem.* **2009**, *109*, 2677–2694.
[95] A. Reyes, F. Moncada, J. Charry, *Int. J. Quantum Chem.* **2019**, *119*, e25705.
[96] F. Pavošević, T. Culpitt, S. Hammes-Schiffer, *Chem. Rev.* **2020**, *120*, 4222–4253.
[97] S. Hammes-Schiffer, *J. Chem. Phys.* **2021**, *155*, 030901.
[98] P. Cassam-Chenaï, B. Suo, W. Liu, *Phys. Rev. A* **2015**, *92*, 012502.
[99] P. Cassam-Chenaï, B. Suo, W. Liu, *Theor. Chem. Acc.* **2017**, *136*, 52.
[100] D. P. O'eill, P. M. W. Gill, *Mol. Phys.* **2005**, *103*, 763–766.
[101] M. Musiał, S. A. Kucharski, *J. Chem. Theory Comput.* **2014**, *10*, 1200–1211.
[102] J. Rumble, Ed. , *CRC Handbook of Chemistry and Physics, 100th Edition*, CRC Press, Boca Raton London New York, **2019**.
[103] C. Pouchan, D. M. Bishop, *Phys. Rev. A* **1984**, *29*, 1–5.
[104] J.-B. Liu, W. H. E. Schwarz, J. Li, *Chem. – Eur. J.* **2013**, *19*, 14758–14767.
[105] H. M. James, *J. Chem. Phys.* **1934**, *2*, 794–810.
[106] H. M. James, *J. Chem. Phys.* **1935**, *3*, 9–14.
[107] C. A. Coulson, W. E. Duncanson, E. N. D. C. Andrade, *Proc. R. Soc. Lond. Ser. Math. Phys. Sci.* **1943**, *181*, 378–386.
[108] E. R. Davidson, S. A. Hagstrom, S. J. Chakravorty, V. M. Umar, C. F. Fischer, *Phys. Rev. A* **1991**, *44*, 7071–7083.
[109] S. Nasiri, M. Zahedi, *Comput. Theor. Chem.* **2017**, *1114*, 106–117.
[110] D. B. Cassidy, A. P. Mills Jr, *Nature* **2007**, *449*, 195–197.
[111] D. B. Cassidy, T. H. Hisakado, H. W. K. Tom, A. P. Mills, *Phys. Rev. Lett.* **2012**, *108*, 133402.
[112] P. Coppens, *X-Ray Charge Densities and Chemical Bonding*, International Union Of Crystallography, Chester, England : Oxford ; New York, **1997**.
[113] R. F. W. Bader, W. H. Henneker, *J. Am. Chem. Soc.* **1965**, *87*, 3063–3068.
[114] R. F. W. Bader, W. H. Henneker, P. E. Cade, *J. Chem. Phys.* **1967**, *46*, 3341–3363.
[115] P. Coppens, *Angew. Chem. Int. Ed. Engl.* **1977**, *16*, 32–40.
[116] J. M. Savariault, M. S. Lehmann, *J. Am. Chem. Soc.* **1980**, *102*, 1298–1303.
[117] W. H. E. Schwarz, P. Valtazanos, K. Ruedenberg, *Theor. Chim. Acta* **1985**, *68*, 471–506.
[118] K. Angermund, K. H. Claus, R. Goddard, C. Krüger, *Angew. Chem. Int. Ed. Engl.* **1985**, *24*, 237–247.
[119] W. H. E. Schwarz, L. Mensching, P. Valtaznos, W. Von Niessen, *Int. J. Quantum Chem.* **1986**, *29*, 909–914.
[120] W. H. E. Schwarz, L. Mensching, P. Valtazanos, W. Von Niessen, *Int. J. Quantum Chem.* **1986**, *30*, 439–444.
[121] K. L. Kunze, M. B. Hall, *J. Am. Chem. Soc.* **1986**, *108*, 5122–5127.
[122] K. L. Kunze, M. B. Hall, *J. Am. Chem. Soc.* **1987**, *109*, 7617–7623.
[123] W. H. E. Schwarz, H. L. Lin, S. Irle, J. E. Niu, *J. Mol. Struct. THEOCHEM* **1992**, *255*, 435–459.
[124] P. Politzer, *Theor. Chim. Acta* **1970**, *16*, 120–125.
[125] W. L. Cao, C. Gatti, P. J. MacDougall, R. F. W. Bader, *Chem. Phys. Lett.* **1987**, *141*, 380–385.
[126] J. Cioslowski, *J. Phys. Chem.* **1990**, *94*, 5496–5498.

[127]  K. E. Edgecombe, J. Vedene H. Smith, F. Müller-Plathe, *Z. Für Naturforschung A* **1993**, *48a*, 127–133.

[128]  L. A. Terrabuio, T. Q. Teodoro, M. G. Rachid, R. L. A. Haiduke, *J. Phys. Chem. A* **2013**, *117*, 10489–10496.

[129]  G. I. Bersuker, C. Peng, J. E. Boggs, *J. Phys. Chem.* **1993**, *97*, 9323–9329.

[130]  A. M. Pendás, M. A. Blanco, A. Costales, P. M. Sánchez, V. Luaña, *Phys. Rev. Lett.* **1999**, *83*, 1930–1933.

[131]  A. Costales, M. A. Blanco, A. Martín Pendás, P. Mori-Sánchez, V. Luaña, *J. Phys. Chem. A* **2004**, *108*, 2794–2801.

[132]  M. Charlton, J. Eades, D. Horváth, R. J. Hughes, C. Zimmermann, *Phys. Rep.* **1994**, *241*, 65–117.

[133]  G. B. Andresen, M. D. Ashkezari, M. Baquero-Ruiz, W. Bertsche, P. D. Bowe, E. Butler, C. L. Cesar, S. Chapman, M. Charlton, A. Deller, S. Eriksson, J. Fajans, T. Friesen, M. C. Fujiwara, D. R. Gill, A. Gutierrez, J. S. Hangst, W. N. Hardy, M. E. Hayden, A. J. Humphries, R. Hydomako, M. J. Jenkins, S. Jonsell, L. V. Jørgensen, L. Kurchaninov, N. Madsen, S. Menary, P. Nolan, K. Olchanski, A. Olin, A. Povilus, P. Pusa, F. Robicheaux, E. Sarid, S. S. el Nasr, D. M. Silveira, C. So, J. W. Storey, R. I. Thompson, D. P. van der Werf, J. S. Wurtele, Y. Yamazaki, *Nature* **2010**, *468*, 673–676.

# Supporting Information

# On the nature of the two-positron bond: Evidence for a novel bond type

Mohammad Goli[1], Dario Bressanini[2] and Shant Shahbazian[3]

[1]School of Nano Science, Institute for Research in Fundamental Sciences (IPM), Tehran 19395-5531, Iran, E-mail: m_goli@ipm.ir

[2]Dipartimento di Scienza e Alta Tecnologia, Università dell'Insubria, Como, Italy
E-mail: dario.bressanini@uninsubria.it

[3]Department of Physics, Shahid Beheshti University, Evin, Tehran 19839-69411, Iran,
E-mail: sh_shahbazian@sbu.ac.ir

# Table of Contents

**Table S1** Total energy of $(PsH)_2$ system as a function of the inter-proton distance computed at the MC-HF/[auc-cc-pVTZ:auc-cc-pVTZ], MC-HF/[auc-cc-pVQZ:auc-cc-pVQZ], VMC and DMC levels of theory for the singlet positronic and electronic spin states. The statistical error of the last digits of the QMC results is given in parentheses. All values are in atomic units.

| $R_{HH}$ | MC-HF | MC-HF | VMC | DMC |
|---|---|---|---|---|
| | aug-cc-pVTZ | aug-cc-pVQZ | | |
| 3.2 | -1.273450 | -1.275147 | | |
| 3.6 | -1.287555 | -1.289003 | | |
| 4.0 | -1.299544 | -1.300855 | | |
| 4.4 | -1.309134 | -1.310377 | | |
| 4.8 | -1.316450 | -1.317666 | | |
| 5.2 | -1.321790 | -1.323002 | | |
| 5.6 | -1.325519 | -1.326738 | -1.58090(2) | -1.5882(1) |
| 5.7 | | | -1.58120(2) | -1.5884(1) |
| 5.8 | | | -1.58149(2) | -1.5886(1) |
| 5.9 | | | -1.58165(2) | -1.5887(1) |
| 6.0 | -1.328008 | -1.329237 | -1.58178(2) | -1.5887(1) |
| 6.1 | | | -1.58180(2) | -1.5887(1) |
| 6.2 | | | -1.58181(2) | -1.5887(1) |
| 6.3 | | | -1.58178(2) | -1.5886(1) |
| 6.4 | -1.329587 | -1.330830 | -1.58166(2) | -1.5884(1) |
| 6.8 | -1.330528 | -1.331792 | | |
| 7.2 | -1.331043 | -1.332332 | | |
| 7.6 | -1.331292 | -1.332603 | | |
| 8.0 | -1.331389 | -1.332710 | | |
| 8.1 | -1.331398 | -1.332720 | | |
| 8.2 | -1.331404 | -1.332725 | | |
| 8.3 | -1.331407 | -1.332727 | | |
| 8.4 | -1.331408 | -1.332725 | | |
| 8.5 | -1.331406 | -1.332720 | | |
| 8.6 | -1.331403 | -1.332713 | | |
| 8.7 | -1.331398 | -1.332705 | | |
| 8.8 | -1.331393 | -1.332696 | | |
| 9.2 | -1.331367 | -1.332651 | | |
| 9.6 | -1.331337 | -1.332607 | | |
| 10.0 | -1.331308 | -1.332568 | | |
| 10.4 | -1.331277 | -1.332537 | | |
| 10.8 | -1.331247 | -1.332512 | | |
| 11.2 | -1.331218 | -1.332490 | | |
| 11.6 | -1.331190 | -1.332470 | | |
| 12.0 | -1.331164 | -1.332453 | | |

**Table S2** Total energy of the singlet electronic spin state of $\left[H^-, e^+, H^-\right]$ system as a function of the inter-proton distance computed at the MC-HF/[auc-cc-pVTZ:auc-cc-pVTZ], VMC and DMC levels of theory. The statistical error of the last digit of the QMC results is given in parentheses. All values are given in atomic units.

| $R_{HH}$ | MC-HF | VMC | DMC |
|---|---|---|---|
| | auc-cc-pVTZ | | |
| 3.2 | -1.111312 | | |
| 3.6 | -1.123523 | | |
| 4.0 | -1.135653 | | |
| 4.4 | -1.146367 | | |
| 4.8 | -1.155217 | | |
| 5.2 | -1.162160 | | |
| 5.6 | -1.167326 | | |
| 5.7 | -1.168361 | | |
| 5.8 | -1.169300 | | |
| 5.9 | -1.170146 | -1.33533(2) | -1.3399(1) |
| 6.0 | -1.170902 | -1.33559(2) | -1.3401(1) |
| 6.1 | -1.171570 | -1.33577(2) | -1.3403(1) |
| 6.2 | -1.172153 | -1.33591(2) | -1.3404(1) |
| 6.3 | -1.172654 | -1.33600(2) | -1.3405(1) |
| 6.4 | -1.173075 | -1.33600(2) | -1.3404(1) |
| 6.5 | -1.173418 | -1.33600(2) | -1.3404(1) |
| 6.6 | -1.173687 | -1.33595(2) | -1.3404(1) |
| 6.7 | -1.173884 | -1.33586(2) | -1.3402(1) |
| 6.8 | -1.174011 | -1.33567(2) | -1.3401(1) |
| 6.9 | -1.174071 | -1.33551(2) | -1.3398(1) |
| 7.0 | -1.174068 | | |
| 7.1 | -1.174003 | | |
| 7.2 | -1.173881 | | |
| 7.6 | -1.172887 | | |
| 8.0 | -1.171278 | | |
| 8.4 | -1.169314 | | |
| 8.8 | -1.167221 | | |
| 9.2 | -1.165157 | | |
| 9.6 | -1.163210 | | |
| 10.0 | -1.161418 | | |
| 10.4 | -1.159786 | | |
| 10.8 | -1.158307 | | |
| 11.2 | -1.156969 | | |
| 11.6 | -1.155761 | | |
| 12.0 | -1.154671 | | |

**Table S3** Vibrational energy levels of $[H^-, e^+, H^-]$ and $(PsH)_2$ based on their respective ground-state potential energy curves obtained at the DMC level of theory given in atomic units.

| Level | $[H^-, e^+, H^-]$ | $(PsH)_2$ |
|---|---|---|
| 0 | -1.3394 | -1.5876 |
| 1 | -1.3373 | -1.5855 |
| 2 | -1.3354 | -1.5836 |
| 3 | -1.3335 | -1.5818 |
| 4 | -1.3316 | -1.5805 |
| 5 | -1.3299 | -1.5793 |
| 6 | -1.3282 | |
| 7 | -1.3267 | |
| 8 | -1.3252 | |
| 9 | -1.3237 | |

**Table S4** Total energy of PsH system computed at the MC-HF/[auc-cc-pVTZ:auc-cc-pVTZ] and MC-HF/[auc-cc-pVQZ:auc-cc-pVQZ] levels of theory while the exact value is from [G. Ryzhikh, J. Mitroy *J. Phys. B* **1999**, 32, 4051]. All values are in atomic units.

|   | MC-HF | MC-HF | Exact |
|---|---|---|---|
|   | auc-cc-pVTZ | auc-cc-pVQZ |   |
| $E$ | -0.665519 | -0.666175 | -0.789196 |

**Table S5** TC-IQA terms computed at the MC-HF/[aug-cc-pVTZ:aug-cc-pVTZ] level of theory at 6.0 Bohr inter-nuclear distance for the singlet positronic and electronic spin states of $(PsH)_2$ system. All values are given in atomic units.

$E_{\text{intra}}$

| $T_e(\Omega_{k=1,2})$ | $V_{\text{e-nuc}}(\Omega_{k=1,2})$ | $V_{ee}^{cl}(\Omega_{k=1,2})$ | $V_{ee}^{xc}(\Omega_{k=1,2})$ | |
|---|---|---|---|---|
| 0.6142 | -1.5609 | 0.9515 | -0.4683 | |
| $T_p(\Omega_{k=1,2})$ | $V_{\text{p-nuc}}(\Omega_{k=1,2})$ | $V_{ep}^{cl}(\Omega_{k=1,2})$ | $V_{pp}^{cl}(\Omega_{k=1,2})$ | $V_{pp}^{xc}(\Omega_{k=1,2})$ |
| 0.0655 | 0.3399 | -0.6081 | 0.1236 | -0.1040 |

$E_{\text{inter}}$

| $V_{\text{e-nuc}}(\Omega_1,\Omega_2)$ | $V_{\text{e-nuc}}(\Omega_2,\Omega_1)$ | $V_{\text{nuc-nuc}}(\Omega_1,\Omega_2)$ | $V_{ee}^{cl}(\Omega_1,\Omega_2)$ | $V_{ee}^{xc}(\Omega_1,\Omega_2)$ | |
|---|---|---|---|---|---|
| -0.3273 | -0.3273 | 0.1667 | 0.6432 | -0.0123 | |
| $V_{ep}^{cl}(\Omega_1,\Omega_2)$ | $V_{ep}^{cl}(\Omega_2,\Omega_1)$ | $V_{\text{p-nuc}}(\Omega_1,\Omega_2)$ | $V_{\text{p-nuc}}(\Omega_2,\Omega_1)$ | $V_{pp}^{cl}(\Omega_1,\Omega_2)$ | $V_{pp}^{xc}(\Omega_1,\Omega_2)$ |
| -0.3050 | -0.3050 | 0.1549 | 0.1549 | 0.1457 | -0.0232 |

**Table S6** TC-IQA terms computed at the MC-HF/[aug-cc-pVTZ:aug-cc-pVTZ] level of theory at 8.4 Bohr inter-nuclear distance for the singlet positronic and electronic spin states of $(PsH)_2$ system. All values are given in atomic units.

$E_{\text{intra}}$

| $T_e(\Omega_{k=1,2})$ | $V_{\text{e-nuc}}(\Omega_{k=1,2})$ | $V_{ee}^{cl}(\Omega_{k=1,2})$ | $V_{ee}^{xc}(\Omega_{k=1,2})$ | |
|---|---|---|---|---|
| 0.6011 | -1.5448 | 0.9368 | -0.4667 | |
| $T_p(\Omega_{k=1,2})$ | $V_{\text{p-nuc}}(\Omega_{k=1,2})$ | $V_{ep}^{cl}(\Omega_{k=1,2})$ | $V_{pp}^{cl}(\Omega_{k=1,2})$ | $V_{pp}^{xc}(\Omega_{k=1,2})$ |
| 0.0652 | 0.3285 | -0.5872 | 0.1182 | -0.1129 |

$E_{\text{inter}}$

| $V_{\text{e-nuc}}(\Omega_1,\Omega_2)$ | $V_{\text{e-nuc}}(\Omega_2,\Omega_1)$ | $V_{\text{nuc-nuc}}(\Omega_1,\Omega_2)$ | $V_{ee}^{cl}(\Omega_1,\Omega_2)$ | $V_{ee}^{xc}(\Omega_1,\Omega_2)$ | |
|---|---|---|---|---|---|
| -0.2373 | -0.2373 | 0.1190 | 0.4730 | -0.0028 | |
| $V_{ep}^{cl}(\Omega_1,\Omega_2)$ | $V_{ep}^{cl}(\Omega_2,\Omega_1)$ | $V_{\text{p-nuc}}(\Omega_1,\Omega_2)$ | $V_{\text{p-nuc}}(\Omega_2,\Omega_1)$ | $V_{pp}^{cl}(\Omega_1,\Omega_2)$ | $V_{pp}^{xc}(\Omega_1,\Omega_2)$ |
| -0.2309 | -0.2309 | 0.1158 | 0.1158 | 0.1129 | -0.0052 |

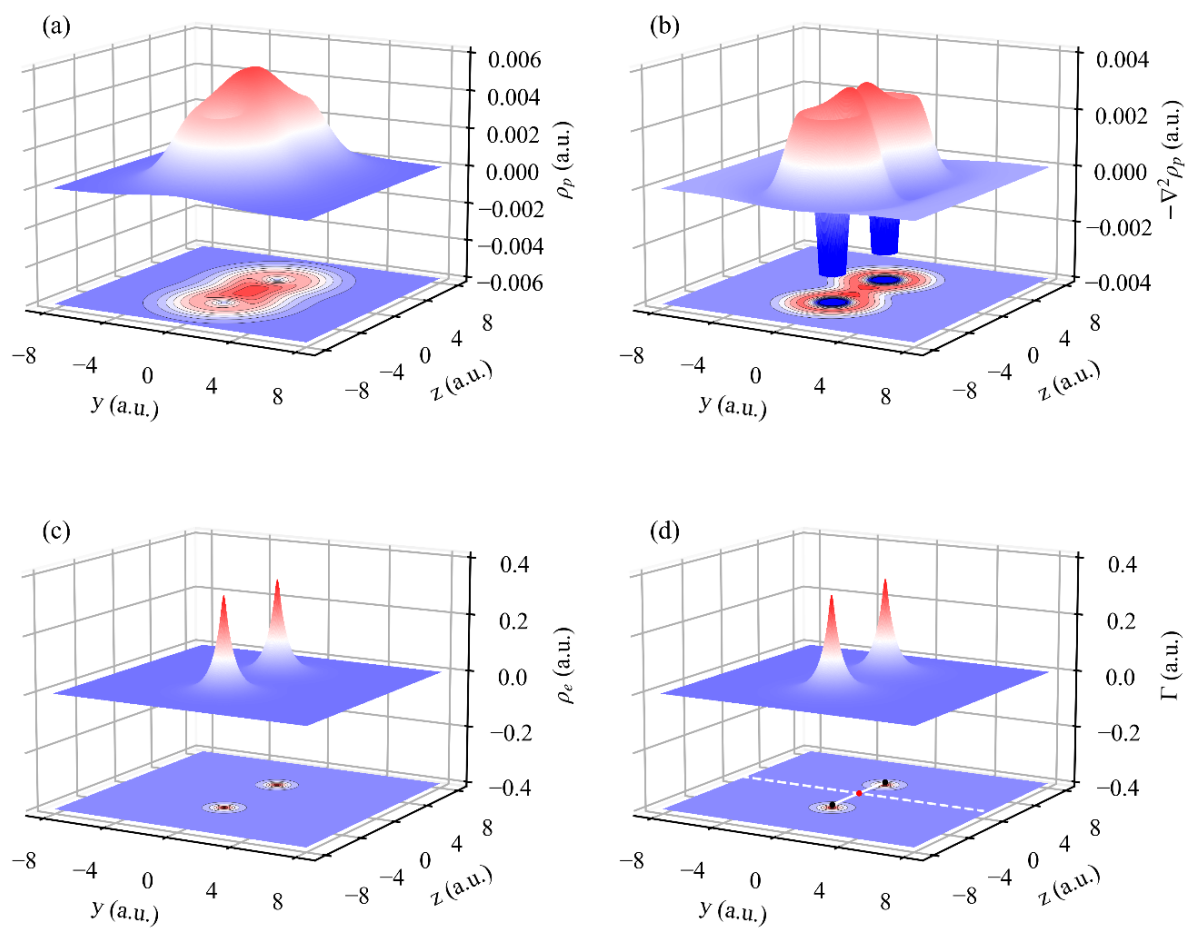

**Fig. S1** Relief maps of (a) the one-positron density, (b) the Laplacian of the one-positron density, (c) the one-electron density, and (d) the Gamma density of $(PsH)_2$. The protons are 6.0 Bohr apart, placed at (0.0, 0.0, 3.0) and (0.0, 0.0, -3.0) in the coordinate system. The black and red spheres in panel (d) are the (3, -3) and (3, -1) CPs of the Gamma density, respectively, while the solid white lines are the gradient paths connecting the (3, -3) and (3, -1) CPs. The dashed white line is the intersection of the zero-flux surface and the yz plane, which acts as the boundary of the AIM.

8